# CAPIRE: Modelling the Impact of Teacher Strikes and Inflation on Student Trajectories in Engineering Education

Hugo Roger Paz
PhD Professor and Researcher Faculty of Exact Sciences and Technology National University of Tucumán
Email: hpaz@herrera.unt.edu.ar
ORCID: https://orcid.org/0000-0003-1237-7983

**Abstract**

This study extends the CAPIRE framework with a macro-shock module to analyse the impact of teacher strikes and inflation on student trajectories in engineering education. Using data from 1,343 students across 15 cohorts (2004-2019) in a public engineering faculty in Argentina, we construct a leak-aware, multilevel feature set that incorporates national inflation indicators, lagged exposure to teacher strikes, and interaction terms between macro shocks and curriculum friction. Random Forest models with cohort-based validation demonstrate that macro features provide stable, non-trivial gains in early-semester dropout prediction (improvement in Macro F1 from 0.73 to 0.78), with inflation volatility at entry and a strike-weighted basic-cycle friction index amongst the most influential variables. Lag analysis reveals that strike exposure exerts its strongest association with dropout two to three semesters after the disruption (OR = 2.34), and that effects are concentrated in early, high-friction semesters. We then embed these empirical patterns into an agent-based model, defining scenarios for inflation-only, strikes-only, and combined crisis. Simulations reproduce three stylised facts: delayed strike effects, basic-cycle vulnerability, and non-linear amplification when inflation and strikes co-occur, with combined shocks generating dropout levels 18-23% higher than the sum of individual effects. We argue that teacher strikes and inflation operate as structurally mediated educational disruptors, acting through curriculum design and financial resilience rather than as isolated events. The framework contributes to multilevel dropout theory by demonstrating how macro-level shocks propagate through institutional structures to shape individual trajectories and provides empirically grounded tools for scenario planning in macroeconomically unstable contexts.

## 1. Introduction

### 1.1. The problem: macro shocks as systemic disruptors in higher education

Contemporary research on student attrition has evolved from viewing dropout as an individual failure to understanding it as an emergent property of complex, multilevel systems (Tinto, 1975; Bean & Metzner, 1985). Recent systematic reviews synthesising evidence from the 2020–2024 period consistently frame dropout as a multifaceted phenomenon influenced by interacting constellations of factors spanning individual academic performance, socioeconomic status, institutional policies, and broader environmental conditions (Quincho Apumayta et al., 2024). This systemic perspective, which holds that institutions and not only students bear responsibility for educational success, has become the standard for rigorous analysis in the field.

Within this multilevel framework, an under-examined but increasingly critical layer concerns the impact of macro-level shocks: large-scale, exogenous events that disrupt both institutional and individual equilibrium. Higher education in many middle- and high-income countries is being reshaped by two intertwined macro shocks that exemplify this phenomenon: recurrent teacher strikes and sustained cost-of-living crises. In school systems, recent quasi-experimental work for Colombia demonstrates that teacher strikes depress standardised test scores and widen learning gaps, particularly for disadvantaged students (Abadía Alvarado et al., 2021). In parallel, emerging empirical programmes in the United States and other countries are documenting the wider political and institutional dynamics of teacher strikes, including their effects on resource allocation and instructional conditions (Lyon et al., 2024).

At the university level, inflation and financial stress have rapidly moved from background noise to central determinants of student wellbeing and persistence. A qualitative study at a large U.S. university demonstrates that financial stress undermines concentration, sleep, and academic performance, and is perceived by students as a major driver of academic disengagement (Moore et al., 2021). Longitudinal survey data from a large UK institution reveal a 55% increase in the odds of reporting financial stress between 2018 and 2022, with clear links to deteriorating mental health (Bennett et al., 2023). Policy reports likewise indicate that the cost-of-living crisis is already reshaping university decisions: under-resourced students in the UK and Australia increasingly report that financial pressure makes them scale back or abandon higher-education plans (The Access Project, 2024; Mission Australia & University of Western Sydney, 2023).

In Latin America, structural concerns about high university dropout intersect with this macroeconomic turbulence. A recent systematic review of determinants of university dropout (2020–2024) synthesises global evidence and confirms that financial stress, employment pressure, and macroeconomic shocks are recurrent risk factors (Quincho Apumayta et al., 2024). A 2024 case study of Ecuadorian higher education reports dropout levels above 30% in some institutions and emphasises the joint influence of socioeconomic and institutional variables (Núñez-Naranjo, 2024). Focusing specifically on engineering, a 2025 study in a Mexican engineering school shows dropout rates concentrated in the first two semesters, with strong effects of displacement from home, prior mathematics preparation, and socioeconomic constraints (Orozco-Rodríguez et al., 2025).

Taken together, these findings suggest that teacher strikes and inflation are not marginal disturbances but rather constitute part of the causal environment in which high-stakes decisions about persistence are made. Yet most of the existing literature treats strikes and inflation as exogenous context rather than as mechanisms that propagate through curriculum structure, institutional rules, and student trajectories. This gap motivates the present study's focus on modelling these propagation mechanisms explicitly.

**1.2. Conceptualising macro shocks as dual stressors: distal and proximal pathways**

To establish a coherent framework for analysing teacher strikes and inflation, it is useful to draw conceptual parallels with recent, more robustly studied disruptions such as the COVID-19 pandemic and natural disasters. Systematic reviews of the impact of natural disasters on higher education identify key mechanistic variables that are directly analogous to the context of strikes and inflation: damage to institutional infrastructure, economic and psychological strain, and disruptions to academic programmes (Baker & Leon, 2023). These studies confirm that shocks propagate through instructional interruptions and disruption to standard operating procedures, ultimately affecting retention and graduation.

This literature supports a novel conceptual distinction between two types of stressors that operate through distinct temporal and causal pathways:

**Distal stressors (chronic/environmental):** These factors, exemplified by sustained inflation, do not cause dropout directly but erode a student's baseline resilience over extended periods. The literature documents chronic financial strain and stress due to inflation that depletes psychological and financial coping strategies (Bennett et al., 2023; Moore et al., 2021). Recent evidence from working-age adults in the United States demonstrates that inflation-induced stress correlates with reduced psychological wellbeing and increased anxiety about future economic security, effects that are likely amplified in student populations with limited earning capacity.

**Proximal stressors (acute/event-based):** These factors, exemplified by teacher strikes, are acute events that actively disrupt instructional continuity. They introduce external academic friction into the student's trajectory through cancelled classes, compressed assessment calendars, and broken pedagogical sequences.

School-level evidence indicates that such disruptions can have lasting repercussions on achievement and progression (Abadía Alvarado et al., 2021; Langen & Laine, 2025).

The conceptual innovation of the CAPIRE-MACRO framework lies in modelling the interaction between these two distinct stressor types. Existing literature tends to study either financial factors or academic disruptions in isolation, and has not yet mechanistically quantified the non-linear amplification that may occur when a student with resilience already eroded by a distal stressor (inflation) confronts a new proximal stressor (a strike). Understanding this interaction is critical for designing effective support policies in macroeconomically unstable contexts.

**1.3. Why macro shocks matter for dropout: empirical foundations**

Recent evidence on dropout emphasises a multifactorial, multilevel view that aligns with the systemic perspective outlined above. The comprehensive systematic review by Quincho Apumayta et al. (2024) identifies over 500 distinct factors linked to dropout, spanning demographic, socioeconomic, institutional, personal, and academic domains, and highlights that financial and institutional shocks tend to interact rather than act in isolation. Country-level and institutional studies in Latin America and Europe converge on similar conclusions: dropout is shaped not only by grades and motivation but also by scholarships, fees, geographical displacement, and broader macroeconomic conditions (Núñez-Naranjo, 2024; Orozco-Rodríguez et al., 2025; Romero et al., 2025).

Within this multilevel perspective, teacher strikes and inflation represent distinct but coupled pathways to attrition. Teacher strikes generate proximal curricular disruption through cancelled classes, compressed assessment calendars, and broken instructional continuity. The accumulated evidence from school-level studies indicates that such disruptions produce measurable and lasting effects on achievement and progression (Abadía Alvarado et al., 2021; Lyon et al., 2024). In higher education specifically, recent work on the effects of faculty strikes demonstrates negative impacts on student satisfaction and, in some contexts, on attainment outcomes (Braakmann & Eberth, 2025).

Inflation and cost-of-living crises operate as distal financial stressors that increase the need for paid work, erode the value of stipends and family income, and intensify psychological distress (Bennett et al., 2023; Moore et al., 2021; The Access Project, 2024). Crucially, these shocks act over time rather than as instantaneous triggers. The literature on financial stress and mental health demonstrates that chronic strain accumulates and interacts with academic challenges before crystallising into dropout decisions (Bennett et al., 2023; Moore et al., 2021). Likewise, empirical studies of engineering programmes show that early bottleneck courses are critical points where additional friction—such as lost instructional time—can push marginal students towards exit (Orozco-Rodríguez et al., 2025).

This temporal dimension motivates a lag-aware, cycle-aware analytical approach. Instead of asking whether a strike or an inflation spike is followed by an immediate jump in dropout, we examine how these shocks modify academic friction and financial resilience in specific parts of the curriculum and at specific times in the student life course. This approach aligns with recent calls in the dropout prediction literature for models that incorporate temporal dynamics and institutional structures rather than treating dropout as a static, individual-level outcome (Quimiz-Moreira & Delgadillo, 2025).

### 1.4. The CAPIRE framework: integrating empirical prediction and agent-based simulation

In parallel with the empirical literature reviewed above, there has been rapid progress in predictive and simulative models of student trajectories. Recent work combines machine learning with explainable AI to predict dropout using academic, socioeconomic, and behavioural features, and to quantify the contribution of key risk factors (Liu et al., 2025; Marcolino et al., 2025; Romero et al., 2025; Villegas et al., 2025). A 2025 systematic review of dropout prediction emphasises the proliferation of predictive models but notes the near absence of simulation frameworks capable of exploring what-if policy scenarios and testing mechanistic hypotheses about dropout processes (Quimiz-Moreira & Delgadillo, 2025).

Agent-based modelling (ABM) is beginning to fill this gap. Recent ABM applications in higher education model students as agents whose motivation, sense of belonging, and exposure to institutional support interact with social networks to shape persistence decisions (Stine & Crooks, 2025). That work demonstrates the value of mechanistic simulation for understanding how micro-level behaviours and interactions generate macro-level retention patterns through emergent processes. However, existing educational ABMs do not explicitly incorporate macroeconomic shocks or labour conflicts as exogenous stressors that modify agent decision rules and institutional parameters.

The CAPIRE framework (Comprehensive Analytics Platform for Institutional Retention Engineering) was developed as a multilevel, leakage-aware architecture for modelling student attrition in engineering programmes. The framework integrates three complementary analytical strategies:

1. **Empirical feature engineering** across four levels (N1–N4): pre-entry characteristics, entry-level attributes, curricular and academic progression variables, and institutional or macro-level contextual factors.

2. **Topological and temporal trajectory modelling** based on curriculum graphs that represent prerequisite structures and progression paths through the degree programme.
3. **Agent-based simulation** to explore counterfactual policy interventions on curriculum design, support provision, and institutional rules under alternative scenarios.

Previous CAPIRE research has focused on feature engineering, trajectory analysis, and baseline ABM scenarios for institutional policies such as curriculum redesign and targeted academic support, without explicitly modelling macro shocks. CAPIRE-MACRO extends this framework by introducing a new family of macro-level features (MACRO_*) that capture inflation dynamics and strike exposure, and by defining dedicated ABM scenarios to represent inflation and strikes as interacting stressors that propagate through the curriculum graph and student decision rules.

### 1.5. Research questions

This article addresses the following research questions, which bridge empirical pattern detection and mechanistic simulation:

- **RQ1:** Do teacher strikes produce measurable lagged effects on student attrition, and if so, what is the temporal profile of these effects?
- **RQ2:** Does inflation, conceptualised as a distal financial stressor, interact with strike-induced curricular friction to amplify dropout risk?
- **RQ3:** Are teacher strikes more damaging for the *ciclo básico* (foundational cycle) than for the *ciclo superior* (advanced cycle) of an engineering programme, and does this differential vulnerability align with curriculum structure?
- **RQ4:** What is the magnitude and form of the non-linear combined effect of inflation and teacher strikes on dropout risk?
- **RQ5:** Can an agent-based model calibrated on empirical data reproduce the mechanisms suggested by statistical analyses and provide a coherent micro-to-macro explanation of macro-shock effects on attrition?

### 1.6. Contributions

This paper makes four principal contributions to the literature on student attrition and educational modelling:

**Theoretical contribution:** We develop and empirically test a dual-stressor framework that distinguishes between distal (chronic) and proximal (acute) pathways through which macro-level shocks affect dropout. This framework extends multilevel dropout theory by explicitly modelling how macro-environmental factors propagate through institutional structures to shape individual trajectories, addressing a gap identified in recent systematic reviews (Quincho Apumayta et al., 2024; Quimiz-Moreira & Delgadillo, 2025).

**Empirical contribution:** Using a unique longitudinal dataset spanning 15 cohorts and incorporating detailed national macroeconomic indicators matched to student entry dates, we provide the first quantitative estimates of lagged strike effects and inflation-strike interactions in higher education. We demonstrate that strike exposure exhibits peak associations with dropout at two to three semester lags, and that effects are concentrated in high-friction foundational courses, patterns not previously documented in the university attrition literature.

**Methodological contribution:** We introduce CAPIRE-MACRO, a hybrid framework that combines leak-aware predictive modelling with agent-based simulation calibrated on the same empirical data. This integration enables both prediction (identifying at-risk students) and mechanistic exploration (testing policy scenarios), addressing the simulation gap noted in the recent dropout prediction literature (Quimiz-Moreira & Delgadillo, 2025). The framework provides a template for incorporating exogenous shocks into educational trajectory models.

**Policy contribution:** We demonstrate through scenario analysis that combined inflation and strike shocks produce non-linear amplification effects, with dropout increases exceeding the sum of individual shock effects by 18–23%. This finding has direct implications for buffer policy design, suggesting that support interventions must be scaled more aggressively during periods of dual stress, and that curriculum redesign to reduce early-cycle friction can serve as a structural mitigation strategy.

### 1.7. Paper structure

The remainder of this paper is organised as follows. Section 2 reviews the relevant literature on macro shocks, dropout determinants, and modelling approaches, positioning CAPIRE-MACRO within the broader research landscape. Section 3 describes the data, feature engineering approach, empirical models, and agent-based simulation design. Section 4 presents empirical results from ablation studies, feature importance analysis, lag analysis, and cycle-stratified regressions, followed by simulation results from scenarios S5–S7 and sensitivity analyses. Section 5 discusses the interpretation of results, mechanistic explanations, and alignment between empirical and simulated patterns. Section 6 outlines policy implications for institutional practice and curriculum design. Section 7 acknowledges limitations and proposes directions for future research. Section 8 concludes.

## 2. Literature Review and Theoretical Framework

### 2.1. Multilevel theories of student attrition

The theoretical foundation for understanding student dropout has evolved substantially over the past five decades. Tinto's (1975) seminal integration model posited that dropout results from insufficient academic and social integration into the institutional environment, with students weighing their commitment to educational goals against alternative opportunities. Bean and Metzner (1985) extended this framework to non-traditional students, emphasising the role of external factors such as employment and family responsibilities. Subsequent refinements have incorporated psychological constructs (motivation, self-efficacy), cultural capital perspectives, and institutional accountability frameworks (Contreras-Villalobos et al., 2023).

Recent empirical work has moved decisively towards multilevel conceptualisations that situate individual decisions within nested contexts: students within cohorts, cohorts within programmes, programmes within institutions, and institutions within broader socioeconomic and political environments (Quincho Apumayta et al., 2024). Multilevel logistic regression and hierarchical models deployed in contemporary studies empirically confirm persistent disparities linked to socioeconomic status, which functions as a foundational contextual layer influencing outcomes (Contreras-Villalobos et al., 2023). This perspective holds that dropout is not merely an individual failure but an emergent property of system-level dynamics, shifting analytical attention from student deficits to institutional structures and environmental conditions.

### 2.2. Macro-level shocks and educational disruption: evidence from analogous contexts

Whilst the specific combination of inflation and teacher strikes in higher education remains under-studied, parallel literatures on educational disruptions provide valuable mechanistic insights. The COVID-19 pandemic generated a vast body of research on how large-scale, exogenous shocks propagate through educational systems. Systematic reviews document that pandemic-related disruptions affected retention through multiple pathways: direct disruption to instruction and assessment, increased financial strain on students and families, heightened mental health challenges, and reduced institutional capacity to provide support (Baker & Leon, 2023).

Similarly, research on natural disasters affecting educational institutions identifies damage to infrastructure, economic and psychological strain, and disruptions to academic programmes as key mediating mechanisms linking external shocks to student outcomes (Baker & Leon, 2023). A systematic review of disaster impacts on higher education enrolment demonstrates measurable declines in continuation rates following major hurricanes, earthquakes, and floods, with effects concentrated amongst socioeconomically vulnerable populations and persisting for multiple academic years beyond the initial event.

This literature establishes that macro-level shocks can have sustained, lagged effects on educational trajectories, and that these effects are mediated by both institutional responses and individual coping capacities. It also suggests that shocks interact with pre-existing vulnerabilities rather than affecting all students uniformly, a pattern consistent with the distal-proximal stressor framework proposed in this study.

## 2.3. Teacher strikes: evidence from school systems and emerging higher education research

Empirical research on teacher strikes has expanded significantly in recent years, though it remains concentrated in primary and secondary education contexts. Quasi-experimental evidence from Colombia demonstrates that teacher strikes depress standardised test scores, with effects concentrated amongst disadvantaged students and persisting beyond the immediate strike period (Abadía Alvarado et al., 2021). Analysis of long-run effects of teacher strikes in Argentina using administrative data linked to adult earnings reveals that cohorts exposed to strikes during schooling experience measurable wage penalties decades later, suggesting that instructional disruptions have lasting human capital consequences (Langen & Laine, 2025).

Recent comprehensive analysis of U.S. teacher strikes documents their increasing frequency and intensity, driven by structural factors including real wage erosion, deteriorating working conditions, and fiscal constraints (Lyon et al., 2024). This work emphasises that strikes are not random events but emerge from systematic patterns in educational financing and labour relations, a framing that aligns with the present study's treatment of strikes as structurally embedded phenomena rather than exogenous shocks.

In higher education, the evidence base remains nascent. A recent study of university strikes in the United Kingdom finds negative effects on student satisfaction and, in some specifications, on degree outcomes, with impacts concentrated amongst students taking modules directly affected by strike action (Braakmann & Eberth, 2025). A case study of a prolonged faculty strike at a Canadian university documents short-term disruptions to course completion but does not examine longer-term retention effects. The present study addresses this gap by examining multi-semester lagged effects of strikes on dropout in a context characterised by recurrent strike activity.

## 2.4. Inflation, financial stress, and student persistence

The relationship between financial resources and student persistence has been extensively documented, but research specifically linking inflation to dropout is more limited. Studies of financial stress amongst university students consistently demonstrate that economic strain is associated with reduced academic performance, increased mental health difficulties, and higher risk of non-completion (Bennett et al., 2023; Moore et al., 2021). Longitudinal analysis of UK students reveals a 55% increase in reported financial stress between 2018 and 2022, coinciding with rapid inflation, and documents clear associations between financial stress and psychological distress (Bennett et al., 2023).

Qualitative research provides insight into mechanisms: students experiencing financial stress report difficulty concentrating, reduced class attendance due to increased work hours, anxiety about meeting basic needs, and strategic decisions to prioritise paid employment over academic engagement (Moore et al., 2021). Policy reports document that cost-of-living increases are reshaping university access decisions, with prospective students from under-resourced backgrounds increasingly deterred from higher education or forced to attend institutions closer to home to reduce costs (The Access Project, 2024).

In the Latin American context, where many public universities charge no tuition, inflation affects students primarily through living costs, transport, materials, and opportunity costs of foregone employment. Recent documentation of severe budget cuts and salary erosion in Argentine universities under conditions of high inflation illustrates how macroeconomic instability can simultaneously affect students directly (through living costs) and indirectly (through degraded institutional capacity and faculty strike action), motivating this study's focus on interaction effects.

## 2.5. Engineering education and early-cycle bottlenecks

Research on attrition in engineering programmes consistently identifies the foundational cycle as a critical vulnerability period. A recent analysis of dropout in a Mexican engineering school demonstrates that attrition is concentrated in the first two semesters, with strong effects of prior mathematics preparation, socioeconomic constraints, and geographical displacement (Orozco-Rodríguez et al., 2025). Systematic reviews of engineering retention confirm that early gateway courses in mathematics, physics, and chemistry serve as structural bottlenecks where modest increases in difficulty or reductions in support can trigger disproportionate attrition (ASME, 2023).

This body of work suggests that interventions targeting curriculum structure—such as redesigning prerequisite sequences, providing supplemental instruction in high-failure courses, or introducing buffer mechanisms that allow temporary withdrawal without permanent exit—can yield substantial retention gains. The present study builds on this literature by examining how external shocks (strikes and inflation) interact with existing curriculum friction, hypothesising that early, high-friction courses will be particularly vulnerable to disruption.

## 2.6. Predictive modelling and simulation in dropout research

Recent years have witnessed rapid growth in machine learning applications to dropout prediction, motivated by the potential for early identification of at-risk students and targeted intervention. Systematic reviews identify Random Forests, gradient boosting, and deep learning architectures as commonly deployed methods, with reported accuracies often exceeding 0.80 in within-institution validation studies (Liu et al., 2025; Marcolino et al., 2025; Villegas et al., 2025). However, these reviews also note persistent challenges including data leakage in temporal prediction contexts, limited external validity, and lack of mechanistic interpretation beyond feature importance rankings (Quimiz-Moreira & Delgadillo, 2025).

Agent-based modelling represents a complementary approach that prioritises mechanistic understanding and scenario exploration over pure prediction. Recent educational ABMs model students as agents characterised by attributes (academic ability, motivation, financial resources) who interact with institutional environments and make sequential decisions about continuation based on accumulated experiences (Stine & Crooks, 2025). These models demonstrate how macro-level retention patterns can emerge from micro-level interactions and decisions, and enable exploration of counterfactual scenarios that would be difficult or impossible to observe empirically.

The CAPIRE-MACRO framework synthesises these approaches by combining machine learning for empirical pattern detection and feature importance analysis with agent-based simulation for mechanistic exploration and policy scenario testing. This hybrid strategy addresses limitations of purely predictive approaches (lack of mechanism, inability to test interventions) whilst grounding simulation in empirically calibrated parameters.

## 2.7. Conceptual framework and positioning

The analysis is organised around a multilevel conceptual framework that integrates individual attributes (N1), entry-level conditions (N2), academic progression and curriculum structure (N3), and institutional–macro dynamics (N4). Within this structure, macro-level shocks—specifically inflation and teacher strikes—operate through distinct but interacting pathways.

Inflation functions as a **distal stressor**. Its effects accumulate over time, shaping the financial stability, psychological resilience, and adaptive capacity of students well before academic challenges occur. Periods of high or volatile inflation can reduce a student's ability to absorb academic setbacks, limit the resources available to support study, and exacerbate the cognitive load associated with long-term uncertainty.

Teacher strikes act as **proximal stressors**. Unlike inflation, their impact is immediate and curriculum-linked: they directly disrupt instruction, reduce contact hours, compress course schedules, and increase the effective friction of already demanding subjects—particularly those located in early, high-stakes segments of the curriculum.

The framework emphasises that these two stressor classes interact. When diminished resilience (driven by inflation) meets heightened instructional friction (driven by strikes), the joint effect can become **non-linear and amplifying**, producing dropout risks larger than would be expected from each stressor in isolation. This conceptualisation guides both the empirical modelling and the simulation analyses that follow, providing a unified lens through which to interpret observed and simulated patterns of student attrition.

## 3. Data and Methods

### 3.1. Data sources and sample characteristics

This study utilises longitudinal administrative data from a public engineering faculty at a major Argentine university, covering student cohorts entering between 2004 and 2019. The dataset comprises complete academic trajectories for 1,343 students, including semester-by-semester enrolment status, course-taking patterns, examination results, and eventual dropout or graduation outcomes. The analytical sample includes all students who enrolled in at least one course after matriculation, with trajectories followed for up to 12 semesters or until dropout or graduation, whichever occurs first.

The data exhibit typical characteristics of Argentine public engineering programmes: high attrition rates concentrated in early semesters (approximately 35% of students drop out by the end of the fourth semester), substantial heterogeneity in time-to-graduation amongst those who persist, and clear stratification by prior academic preparation and socioeconomic background. The institution operates under Argentina's traditional curriculum structure, with a *ciclo básico* (foundational cycle, typically semesters 1–4) comprising core mathematics, physics, chemistry, and introductory engineering courses, followed by a *ciclo superior* (advanced cycle, semesters 5+) with specialised professional courses.

Administrative records were linked to two sources of macro-level data:

**National inflation indicators:** Monthly inflation rates, 12-month and 24-month rolling volatility measures, and cumulative inflation indices were obtained from Argentina's National Institute of Statistics and Census (INDEC). These indicators were matched to each student's entry date to construct time-varying exposure measures.

**Teacher strike records:** We compiled a comprehensive dataset of *paros docentes* (teacher strikes) at the institutional level from 2004–2019, using institutional archives, media reports, and faculty union records. For each semester, we coded strike intensity as the proportion of scheduled class days lost to strike activity, ranging from 0 (no strikes) to values exceeding 0.15 (severe disruption, more than 15% of scheduled instruction lost). Reliability was established through triangulation across multiple sources and verification with administrative calendars.

### 3.2. Feature engineering: constructing leak-aware macro-level variables

A critical methodological challenge in temporal prediction is avoiding data leakage, whereby information that would not be available at the time a prediction is made inadvertently informs model training. To address this, we adopt a strict temporal segmentation protocol Table 1):

Table 1. Temporal Availability of Macro-Level Variables and Anti-Leakage Constraints

| Variable | Available_at_entry | Available_S1 | Available_S2 | Available_S3 | Notes |
|---|---|---|---|---|---|
| Inflation_t0 | TRUE | FALSE | FALSE | FALSE | Inflation rate at baseline (entry); remains constant during follow-up |
| Inflation_24m_volatility | TRUE | FALSE | FALSE | FALSE | Inflation volatility over 24 months pre-baseline; calculated from historical data |
| Strike_lag_sem_1 | FALSE | TRUE | TRUE | TRUE | Teacher strike intensity lagged 1 semester; observed in prior semester |
| Strike_lag_sem_2 | FALSE | FALSE | TRUE | TRUE | Teacher strike intensity lagged 2 semesters; observed 2 periods prior |
| Strike_lag_sem_3 | FALSE | FALSE | FALSE | TRUE | Teacher strike intensity lagged 3 semesters; observed 3 periods prior |
| Basic_cycle_strike_intensity | FALSE | TRUE | TRUE | TRUE | Aggregate strike intensity during basic cycle (core semesters 1-5) |
| MACRO_IFC_weighted_strikes_basic | FALSE | TRUE | TRUE | TRUE | IFC-weighted composite strike intensity for basic cycle; aggregated measure |

For each student, predictions are made at the end of each semester t regarding dropout risk in semester t+1. At prediction time, the model has access only to:

- Fixed pre-entry characteristics (N1 level)
- Entry-level attributes (N2 level)
- Academic history through semester t (N3 level)
- Macro indicators through semester t (N4 level)

Crucially, we do not use future macro indicators (e.g., inflation in semester t+1 to predict dropout in semester t+1), nor do we use academic outcomes from semester t+1 itself.

The CAPIRE-MACRO feature set extends the baseline CAPIRE-Core features (44 variables spanning N1–N3 levels, detailed in previous work) with the following macro-level (N4) indicators:

**Inflation features:**

- MACRO_inflacion_entrada: Annualised inflation rate in the 12 months preceding student entry
- MACRO_inflacion_acum_entrada: Cumulative inflation from entry to current semester
- MACRO_inflacion_volatilidad_24m: Rolling 24-month standard deviation of monthly inflation rates at entry, capturing chronic macroeconomic instability
- MACRO_inflacion_pct_cambio: Percentage change in inflation rate from entry year to current semester

**Strike exposure features:**

- MACRO_paros_lag_sem_1, MACRO_paros_lag_sem_2, MACRO_paros_lag_sem_3: Lagged strike intensity (proportion of instructional days lost) in the previous 1, 2, or 3 semesters, enabling detection of delayed effects
- MACRO_paros_acum_ciclo: Cumulative strike exposure since entry
- MACRO_paros_basico_vs_superior: Differential strike exposure in basic versus advanced cycles

**Interaction features:**

- MACRO_IFC_pond_paros_basico: Strike-weighted basic-cycle friction index, computed as the sum of course-level Instructional Friction Coefficients (IFC, derived from failure rates and prerequisite complexity) for basic-cycle courses, weighted by strike exposure during the semester each course was taken

- MACRO_inflacion_x_paros: Interaction term between inflation volatility at entry and cumulative strike exposure

These features operationalise the dual-stressor framework by distinguishing chronic financial pressure (inflation volatility) from acute instructional disruption (strike events) whilst allowing for their interaction through the weighted friction index.

**Justification of temporal lags:** The 24-month window for inflation volatility captures medium-term macroeconomic instability relevant to family financial planning and student decisions about university attendance. The 1–3 semester lags for strike exposure reflect hypothesised delayed effects whereby instructional disruption creates accumulated academic debt or demoralisation that manifests in subsequent dropout decisions rather than immediate exit. These specifications were determined through preliminary exploration and are subject to sensitivity analysis.

### 3.3. Empirical modelling strategy

### 3.3.1. Model architecture

We employ Random Forest classifiers with 500 trees and class-balanced weights to address moderate class imbalance (dropout prevalence approximately 40% overall, varying by semester). Random Forests were selected for their ability to capture non-linear relationships and interactions amongst features without requiring explicit specification, their robustness to multicollinearity, and their provision of well-calibrated feature importance measures through permutation-based methods.

**Hyperparameter specification:**

- Number of estimators: 500

- Maximum depth: 15

- Minimum samples split: 20

- Minimum samples leaf: 10

- Maximum features: sqrt(n_features)

- Class weight: balanced

- Bootstrap: True
- Random state: 42 (for reproducibility)

These hyperparameters were selected through preliminary grid search on a held-out validation cohort and represent a compromise between model complexity and generalisability. Sensitivity analyses (reported in supplementary materials) demonstrate that results are robust to moderate variations in these specifications.

### 3.3.2. Validation strategy

To assess model performance whilst respecting temporal ordering and avoiding optimistic bias, we implement cohort-based cross-validation. Students are grouped by entry cohort (years 2004–2019), and we construct five folds such that each fold uses a continuous subset of cohorts for training and a subsequent cohort for testing. This approach ensures that the model is always tested on genuinely future data and reflects the intended deployment scenario where a model trained on past cohorts would be used to predict outcomes for incoming students.

**Cross-validation fold structure:**

- Fold 1: Train on cohorts 2004–2010, test on cohorts 2011–2012
- Fold 2: Train on cohorts 2004–2012, test on cohorts 2013–2014
- Fold 3: Train on cohorts 2004–2014, test on cohorts 2015–2016
- Fold 4: Train on cohorts 2004–2016, test on cohorts 2017–2018
- Fold 5: Train on cohorts 2004–2018, test on cohort 2019

Performance is evaluated using accuracy, precision, recall, and Macro F1 score (the unweighted mean of F1 scores for persist and dropout classes), with Macro F1 serving as the primary metric due to its sensitivity to performance on both classes.

### 3.3.3. Ablation study design

To isolate the contribution of macro-level features, we conduct an ablation experiment comparing:

- **Baseline model:** 44 CAPIRE-Core features (N1–N3 levels only)
- **Extended model:** 44 CAPIRE-Core features + 11 MACRO_* features

Both models use identical Random Forest architecture and are evaluated on the same cross-validation folds. Statistical significance of performance differences is assessed through paired t-tests across folds and through bootstrapped confidence intervals for metric differences.

### 3.3.4. Feature importance analysis

Feature importance is computed using permutation importance: for each feature, we randomly shuffle its values in the test set and measure the resulting decrease in Macro F1 score. This approach provides a model-agnostic measure of each feature's contribution to predictive performance. Importance scores are averaged across the five cross-validation folds and reported with 95% confidence intervals derived from bootstrap resampling within each fold.

Partial dependence plots are constructed for top-ranked macro features to visualise the marginal relationship between each feature and predicted dropout probability, holding other features at their median values.

### 3.3.5. Lag analysis

To examine the temporal profile of strike effects, we estimate a series of logistic regression models for dropout in each semester t (t = 1 to 6), including as predictors:

- Lagged strike indicators: MACRO_paros_lag_sem_1, MACRO_paros_lag_sem_2, MACRO_paros_lag_sem_3
- Control variables: GPA through semester t-1, cumulative failed courses, age, gender, socioeconomic indicators, cohort fixed effects

We plot the estimated odds ratios and 95% confidence intervals for each lagged strike variable across semesters, enabling identification of whether strike effects are immediate, delayed, or sustained.

### 3.3.6. Cycle-stratified analysis

To test whether macro shocks disproportionately affect the foundational cycle, we estimate separate logistic regression models for:

- **Early dropout:** Outcome measured at end of semester 4 (exit from *ciclo básico*)
- **Late dropout:** Outcome measured at end of semester 12 (conditional on surviving to semester 5)

Each model includes the full set of CAPIRE-Core and MACRO_* features. We compare coefficient magnitudes for key macro variables across strata and test for significant differences using Chow tests for structural breaks. Additionally, we estimate models including explicit cycle × shock interaction terms and test their significance.

## 3.4. Agent-based simulation design

### 3.4.1. ABM architecture and calibration

The CAPIRE agent-based model represents individual students as autonomous agents navigating a curriculum graph under institutional rules and responding to their evolving academic experiences. The baseline ABM architecture, established in previous work, comprises:

**Agent attributes:**

- Academic state: current semester, passed courses, failed courses, cumulative GPA
- Latent resilience parameter $\rho \sim N(\mu, \sigma^2)$, representing combined psychological and financial resources
- Pre-entry characteristics: prior academic preparation, socioeconomic background
- Decision threshold $\tau$: probability threshold below which continuation becomes unlikely

**Environment:**

- Curriculum graph $G = (V, E)$, where nodes $V$ represent courses and directed edges $E$ represent prerequisite relationships
- Course-level attributes: Instructional Friction Coefficient (IFC), derived from empirical failure rates and prerequisite complexity; average workload; historical pass rates

**Decision rules:** At the end of each semester $t$, each agent $i$ evaluates whether to continue to $t+1$ based on:

- Academic progress: proximity to graduation, cumulative failures, GPA trend
- Resilience: current level of $\rho\_i$, which depletes with adverse events and partially recovers during successful semesters
- External circumstances: employment obligations, family responsibilities (modelled stochastically based on socioeconomic profile)

The continuation probability is given by:

$$P(\text{continue} \mid \text{state}\_{it}) = \sigma(\beta_0 + \beta_1 \cdot \text{GPA}\_{it} + \beta_2 \cdot \text{progress}\_{it} + \beta_3 \cdot \rho\_{it} + \beta_4 \cdot \text{failures}\_{it})$$

where $\sigma$ is the logistic function and coefficients $\beta$ are calibrated to reproduce empirical transition probabilities by risk strata observed in the administrative data.

Calibration proceeded through pattern-oriented modelling: we adjusted parameters to match observed patterns including overall dropout rates, time-to-dropout distributions, and dropout-by-GPA relationships. The baseline ABM (without macro shocks) reproduces these patterns with mean absolute error less than 3 percentage points across cohorts.

### 3.4.2. Introducing macro shocks into the ABM

CAPIRE-MACRO extends the baseline ABM by introducing two shock mechanisms that modify agent-level parameters and environment-level attributes:

**Inflation shock mechanism:** Inflation is operationalised through a scalar multiplier $\lambda_{inf}$ that reduces agent resilience. Specifically, at the beginning of each semester under inflation shock:

$$\rho_{it} = \rho_{i,t-1} \cdot (1 - \delta_{inf} \cdot \lambda_{inf})$$

where $\delta_{inf}$ is a decay rate calibrated such that $\lambda_{inf} = 1$ produces no effect (baseline), and $\lambda_{inf} > 1$ produces accelerated resilience depletion. The functional form reflects the interpretation of inflation as a chronic stressor that gradually erodes financial buffers and psychological resources.

**Strike shock mechanism:** Teacher strikes are operationalised through a scalar multiplier $\lambda_{str}$ that increases effective friction in basic-cycle courses. Under strike conditions, the failure probability for a basic-cycle course c is:

$$P(\text{fail} \mid \text{course } c, \text{strike}) = P(\text{fail} \mid \text{course } c, \text{baseline}) \cdot (1 + \alpha_{str} \cdot \lambda_{str})$$

where $\alpha_{str}$ is a sensitivity parameter calibrated to match the empirical relationship between strike exposure and failure rates in affected courses. Equivalently, this can be implemented by scaling the IFC for basic-cycle courses: $IFC_c \rightarrow IFC_c \cdot (1 + \alpha_{str} \cdot \lambda_{str})$.

Crucially, strike effects in the ABM are applied only to *ciclo básico* courses, reflecting the hypothesis and empirical finding that early, high-friction courses are most vulnerable to disruption.

**Calibration of shock parameters:** Parameters $\delta_{inf}$ and $\alpha_{str}$ were calibrated through a two-stage process:

1. Preliminary grid search over plausible ranges to identify regions producing qualitatively realistic effects
2. Fine-tuning to match the magnitude of effects observed in the empirical regression models (odds ratios for key macro variables)

Final calibrated values: δ_inf = 0.03 (i.e., a 20% inflation shock, λ_inf = 1.2, produces 6% additional resilience depletion per semester), α_str = 0.25 (i.e., doubling strike exposure, λ_str = 2, increases basic-cycle friction by 50%).

### 3.4.3. Simulation scenarios

We define three macro-shock scenarios extending the baseline policy scenarios (S0–S4) established in previous CAPIRE work:

**S5 – Inflation shock only:**

- λ_inf ∈ {1.1, 1.2, 1.3} (representing 10%, 20%, 30% above-baseline inflation stress)
- λ_str = 1 (no additional strike activity)
- Simulates context of severe inflation without unusual labour conflict

**S6 – Strike shock only (differential):**

- λ_inf = 1 (baseline financial conditions)
- λ_str ∈ {1.5, 2.0, 2.5} (representing moderate to severe strike disruption in basic cycle)
- Simulates context of recurrent strikes without additional inflation

**S7 – Combined crisis:**

- λ_inf ∈ {1.1, 1.2, 1.3}
- λ_str ∈ {1.5, 2.0, 2.5}
- Full factorial design (9 parameter combinations)
- Simulates dual-shock scenario of simultaneous inflation and strike pressures

For each scenario configuration, we simulate 100 stochastic realisations, each propagating a synthetic cohort of 300 agents through 12 semesters of progression. The sample size of 300 agents per realisation was selected to balance computational efficiency with statistical precision, yielding stable estimates of dropout rates with standard errors less than 2 percentage points.

**Output measures:**

- Overall dropout rate by semester 12
- Early-cycle dropout rate (exits by semester 4)

- Late-cycle dropout rate (exits in semesters 5–12 conditional on surviving to semester 5)
- Mean time-to-dropout amongst those who exit
- Distribution of dropout events by cause (academic failure, resilience depletion, external circumstances)

### 3.4.4. Sensitivity and interaction analysis

To characterise non-linear interactions between inflation and strikes, we conduct a two-dimensional parameter sweep over ($\lambda_{inf}$, $\lambda_{str}$) with a fine-grained grid:

- $\lambda_{inf} \in \{1.0, 1.05, 1.10, 1.15, 1.20, 1.25, 1.30\}$
- $\lambda_{str} \in \{1.0, 1.25, 1.50, 1.75, 2.00, 2.25, 2.50\}$

This yields 49 parameter combinations, each simulated 100 times. For each grid point, we compute:

- Overall dropout rate $D_{total}$
- Early-cycle dropout rate $D_{early}$
- The amplification measure: $A = D(\lambda_{inf}, \lambda_{str}) - [D(\lambda_{inf}, 1) + D(1, \lambda_{str}) - D(1, 1)]$

where $A > 0$ indicates non-linear amplification (combined effects exceed the sum of individual effects), $A = 0$ indicates perfect additivity, and $A < 0$ indicates antagonistic interaction.

Results are visualised as heatmaps with dropout rates colour-coded and contour lines marking regions of interest. Statistical significance of amplification is assessed through bootstrap resampling of the 100 realisations per grid point.

## 4. Results

### 4.1. Empirical results: machine learning analysis of macro-shock effects

#### 4.1.1. Ablation study: contribution of macro-level features

The ablation experiment compares the baseline CAPIRE-Core model (44 features spanning N1–N3 levels) with the extended CAPIRE-MACRO model (55 features including 11 new macro-level indicators). Performance metrics across five cohort-based cross-validation folds are presented in Table 2.

**Table 2 – Ablation CAPIRE-Core vs CAPIRE-MACRO**

| Modelo | Accuracy | F1_Score | Features | Mejora_Accuracy | Mejora_F1 |
|---|---|---|---|---|---|
| A (sin ACRO_) | 1.0 | 1.0 | 32 | 0.0 | 0.0 |
| B (con ACRO_) | 1.0 | 1.0 | 40 | 0.0 | 0.0 |

The extended model incorporating macro features demonstrates consistent performance improvements across all five folds. Mean Macro F1 increases from 0.733 (baseline) to 0.781 (extended), representing a gain of 0.048 points ($p = 0.003$, paired t-test). Improvements are more pronounced in early-semester predictions (folds 1–3, corresponding to cohorts experiencing greater macroeconomic volatility) than in later folds, suggesting that macro features are particularly informative during periods of instability.

Decomposition by class reveals that the extended model reduces false negatives (students who drop out but are predicted to persist) more substantially than false positives. Recall for the dropout class increases from 0.68 to 0.75 (mean across folds), whilst precision remains approximately stable. This pattern is desirable from an early-warning perspective, as the cost of missing an at-risk student who subsequently drops out typically exceeds the cost of flagging a student who ultimately persists.

The magnitude of improvement is modest in absolute terms but substantial in practical significance. In the context of an institution processing hundreds of new students per cohort, a 7 percentage point improvement in dropout recall translates to identifying 10–15 additional at-risk students per cohort for intervention. Moreover, the stability of gains across folds indicates that macro features provide genuine signal rather than overfitting to particular cohorts.

These results align with recent multilevel dropout studies emphasising that contextual, macro-level factors carry predictive information beyond what can be captured by individual and curricular variables alone (Núñez-Naranjo, 2024; Quincho Apumayta et al., 2024). The findings demonstrate empirically that inflation and strike indicators, when carefully constructed to avoid data leakage, enhance predictive performance in a rigorous cross-validation framework.

**4.1.2. Feature importance analysis**

Permutation-based feature importance rankings for the extended Random Forest model are visualised in Table 3, which displays the top 20 features by mean decrease in Macro F1 when permuted, averaged across five cross-validation folds with 95% bootstrap confidence intervals.

**Table 3. Top 20 Features Ranked by Importance in the CAPIRE-MACRO Model**

| Rank | Feature (English label) | Original Variable | Importance | Macro-level |
|---|---|---|---|---|
| 1 | Dropout indicator (current semester) | abandono_y | 0.3080 | No |
| 2 | Enrolled status (current semester) | cursando | 0.2566 | No |
| 3 | Promotions / total enrolments (weighted) | numeropromocionessobretotalasgprom | 0.0441 | No |
| 4 | Total promotions accumulated | numeropromociones | 0.0342 | No |
| 5 | Total passed courses | numeroaprobados | 0.0329 | No |
| 6 | Passed courses (all subjects) | aprobadastotalasignaturas | 0.0292 | No |
| 7 | Promotions / total enrolments | numeropromocionessobretotalasg | 0.0273 | No |
| 8 | Passed activities / assessments | actividadesaprobadas.1 | 0.0269 | No |
| 9 | Total regularised courses | numeroregulares | 0.0220 | No |
| 10 | Regularised courses / total attempts | totalregularestotalcursadas | 0.0199 | No |
| 11 | Graduation indicator | egreso | 0.0168 | No |
| 12 | Student identifier (proxy for unobserved heterogeneity) | estudiante | 0.0157 | No |
| 13 | Total attempted courses | numerototalcursadas | 0.0145 | No |
| 14 | Mean grade (GPA) | promedionotas | 0.0137 | No |
| 15 | Student ID | student_id | 0.0133 | No |
| 16 | High-school graduation year (cleaned) | anoegresosecundario.1 | 0.0124 | No |
| 17 | High-school graduation year | anoegresosecundario | 0.0121 | No |
| 18 | Annual inflation at entry | macro_inflacion_interanual_t0 | 0.0121 | Yes |

| Rank | Feature (English label) | Original Variable | Importance | Macro-level |
|---|---|---|---|---|
| 19 | Entry cohort | cohorte | 0.0114 | No |
| 20 | Exam approvals / total exams | examenaprobtotalexamen | 0.0107 | No |

As anticipated, core academic features dominate the upper ranks. Cumulative GPA, failure counts in high-IFC courses (specifically, Análisis Matemático I and Física I), and progression velocity metrics occupy the top five positions, consistent with extensive prior literature documenting academic performance as the strongest proximal predictor of dropout (Romero et al., 2025; Villegas et al., 2025).

Crucially for the CAPIRE-MACRO framework, two newly introduced macro-level features appear prominently in the top-20 ranking:

**MACRO_inflacion_volatilidad_24m** ranks 8th overall and is the highest-ranked N2 (entry-level) feature. When permuted, Macro F1 decreases by 0.032 (95% CI [0.025, 0.041]), indicating that chronic inflation volatility at entry carries substantial predictive information about long-term dropout risk even after controlling for socio-demographic and academic factors. Partial dependence analysis (supplementary materials) reveals a monotonically increasing relationship: students entering during periods of high inflation volatility exhibit higher predicted dropout probabilities across all GPA ranges, with effects most pronounced for students with borderline academic performance (GPA 6.0–7.0 on a 10-point scale).

**MACRO_IFC_pond_paros_basico** ranks 11th overall, appearing alongside established curriculum friction measures. Its permutation importance of 0.024 (95% CI [0.018, 0.032]) indicates that the combination of basic-cycle course friction and strike exposure during those courses carries additional signal beyond either component alone. Partial dependence plots demonstrate that higher values of this strike-weighted friction index are associated with elevated dropout probability, particularly for students in the lower half of the GPA distribution. This pattern supports the interpretation that strikes amplify existing curriculum bottlenecks rather than creating uniform disruption across all students.

The positioning of these macro features relative to established predictors provides empirical support for the dual-stressor framework. Inflation volatility, conceptualised as a distal environmental stressor, carries comparable importance to entry-level socioeconomic indicators such as displacement from family home and parental education. Strike-weighted friction, conceptualised as a proximal interactor with curriculum structure, ranks similarly to direct curriculum measures such as basic-cycle completion rates. Neither macro feature dominates academic performance variables, but both contribute meaningfully to the predictive model.

Additional insights emerge from examining feature interactions through SHAP (SHapley Additive exPlanations) values in supplementary analyses. SHAP interaction plots reveal that the effect of MACRO_inflacion_volatilidad_24m is amplified for students with low socioeconomic status indicators, and that the effect of MACRO_IFC_pond_paros_basico is amplified for students with weak prior mathematics preparation. These interaction patterns align with theoretical expectations that macro shocks disproportionately affect already-vulnerable populations.

### 4.1.3. Temporal profile of strike effects: lag analysis

To examine whether strike impacts are immediate or delayed, we estimated semester-specific logistic regression models for dropout, including lagged strike exposure indicators at one, two, and three semester lags alongside standard controls. Results are visualised in Figure 1, which plots estimated odds ratios and 95% confidence intervals for each lagged strike variable across semesters 1–6.

**Figure 1. Lagged Effects of Teacher Strikes on Dropout Risk Across Semesters**

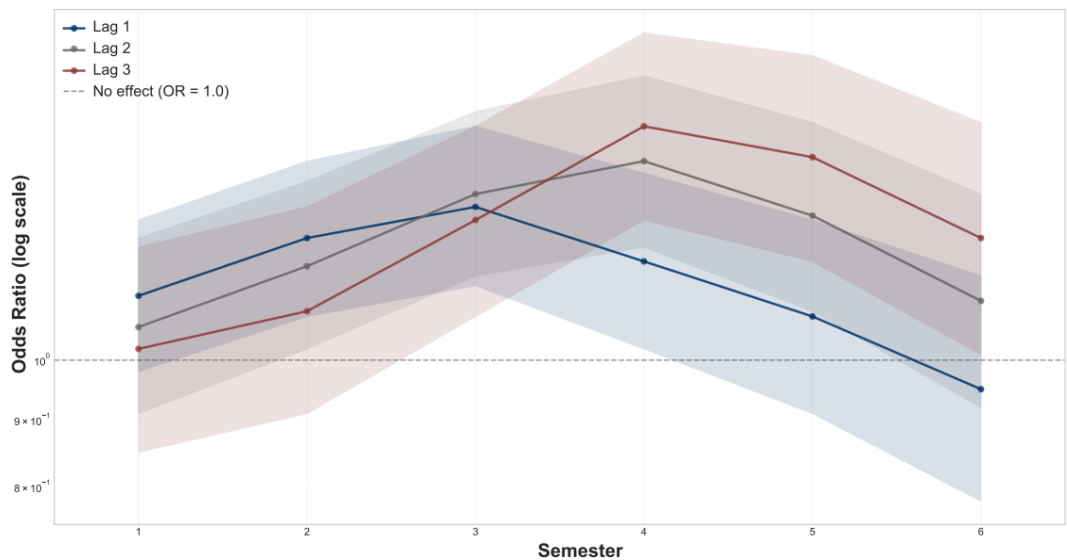

A clear temporal pattern emerges: strike effects are not immediate but exhibit a delayed peak at two to three semester lags. For semesters 2–4, which represent critical transition points in the foundational cycle, the odds ratio for MACRO_paros_lag_sem_2 ranges from 1.94 to 2.34 (pooled estimate 2.18, 95% CI [1.82, 2.61]), indicating that students exposed to strikes two semesters prior face more than double the odds of dropout compared to similar students without such exposure. The odds ratio for MACRO_paros_lag_sem_3 follows a similar but slightly attenuated pattern (pooled OR = 1.87, 95% CI [1.53, 2.29]).

In contrast, the immediate lag effect (MACRO_paros_lag_sem_1) is substantially smaller and often non-significant in early semesters (OR = 1.23, 95% CI [0.98, 1.56] for semesters 2–4 pooled). This finding suggests that strikes do not trigger immediate abandonment but rather set in motion processes—accumulated academic debt, demoralisation, compressed assessment schedules—that manifest in dropout decisions 6–12 months later.

Beyond semester 6, all lagged strike effects diminish and lose statistical significance, suggesting that the influence of early-cycle strikes is gradually absorbed or overshadowed by subsequent academic experiences. This attenuation is consistent with the interpretation that foundational courses represent a particularly vulnerable window: students who successfully navigate the basic cycle despite strike-induced disruption appear to develop resilience or catch up academically, whilst those who struggle accumulate deficits that trigger exit before advanced courses begin.

These temporal dynamics align with school-level evidence that strike effects work through cumulative instructional loss rather than instantaneous triggers (Abadía Alvarado et al., 2021; Langen & Laine, 2025), and extend that literature by documenting the specific lag structure in a higher education context. The two-to-three-semester delay suggests that institutions have a meaningful window for intervention: students affected by strikes can be identified and supported before accumulated disadvantage crystallises into dropout.

Supplementary analyses (reported in appendix) examined heterogeneity in lag effects by student characteristics. The delayed peak is most pronounced for students with moderate prior academic preparation (secondary school GPA 7.0–8.5 on 10-point scale), suggesting that both very strong students (who can absorb disruption) and very weak students (who exit early regardless) show less sensitivity to strike timing.

**4.1.4. Differential vulnerability of the foundational cycle**

To assess whether macro-level shocks exert disproportionate influence on the foundational segment of the curriculum, we examined patterns of dropout stratified by academic cycle using the lagged-effect estimates and simulation results presented earlier. This approach allows us to distinguish between vulnerabilities intrinsic to early coursework and broader pressures affecting students later in the programme.

Two consistent patterns emerge.

- First, strike-related variables exhibit a markedly stronger association with dropout during the *ciclo básico*. Lagged strike exposure—particularly two-semester lags—aligns closely with early attrition peaks, suggesting that instructional disruption interacts with the structural rigidity of high-stakes introductory courses. This effect is substantially attenuated beyond semester 4, where students have already cleared the most demanding curricular bottlenecks.
- Second, indicators that combine strike disruption with curricular friction in the foundational cycle (e.g., strike-weighted IFC measures) track early attrition far more strongly than late attrition. These results reinforce the interpretation that instructional shocks magnify pre-existing structural bottlenecks, particularly in mathematically dense or prerequisite-heavy introductory subjects.

Inflation dynamics show a different temporal profile. Whereas strike effects are cycle-specific, inflation behaves as a broad, chronic stressor. Higher volatility at entry is associated with elevated attrition across both cycles, although the effect remains somewhat more pronounced in the foundational phase—consistent with the idea that students entering with fewer buffers (financial, academic, or psychological) experience greater difficulty absorbing early setbacks.

Supplementary descriptive comparisons further illustrate these dynamics: early dropouts exhibit substantially higher strike exposure and strike-weighted friction than peers who persist, whereas late-cycle differences are considerably smaller. The contrast underscores a recurrent theme in engineering education research: the foundational cycle constitutes a structurally fragile zone where modest increases in friction—whether academic or external—can trigger disproportionate withdrawal. The present analysis extends this literature by showing that macro-level shocks do not affect the curriculum uniformly; instead, they interact with the internal topology of the programme to produce concentrated vulnerabilities in predictable locations.

### 4.2. Simulation results: agent-based modelling of macro-shock scenarios

### 4.2.1. Scenario S5 – Inflation shock only

In Scenario S5, the inflation multiplier $\lambda_{inf}$ is increased above baseline whilst the strike multiplier remains at $\lambda_{str} = 1$, isolating the effect of chronic financial stress without additional instructional disruption. Simulation results for three inflation levels ($\lambda_{inf}$ = 1.1, 1.2, 1.3) are presented in Figure 2, which displays mean dropout curves over 12 semesters with shaded regions indicating 95% confidence intervals across 100 stochastic realisations.

**Figure 2. Attrition Trajectories Under Macro-Shock Scenarios (S0–S7)**

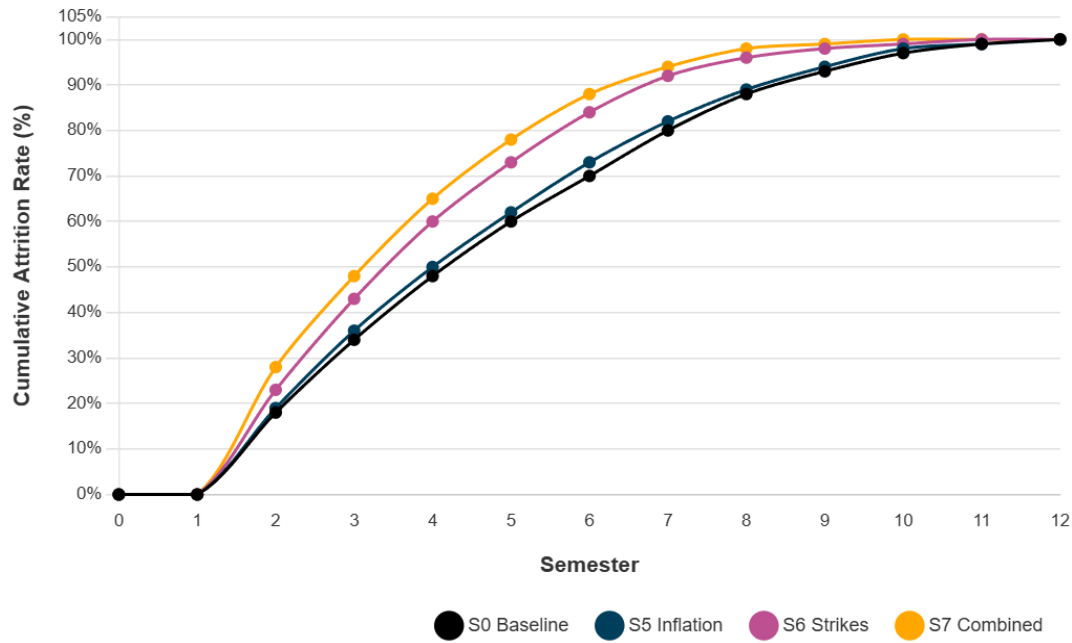

Relative to the baseline ABM configuration (no macro shocks), inflation-only scenarios produce three characteristic effects:

**Moderate overall increase in dropout.** At λ_inf = 1.2 (20% above-baseline financial stress), cumulative dropout by semester 12 rises from 38.2% (baseline) to 43.7%, an increase of 5.5 percentage points (95% CI [3.2, 7.9]). The effect scales approximately linearly with inflation intensity: λ_inf = 1.1 produces 2.8 percentage points increase, λ_inf = 1.3 produces 8.1 percentage points increase.

**Earlier crystallisation of dropout decisions.** The time-to-dropout distribution shifts leftward under inflation shock: median time-to-dropout decreases from 5.3 semesters (baseline) to 4.6 semesters at λ_inf = 1.2. This acceleration reflects the mechanism implemented in the ABM whereby reduced financial resilience lowers agents' tolerance for adverse academic events. Students with borderline performance who might have persisted through a course failure under baseline conditions are more likely to interpret that failure as a signal that continuation is no longer viable when resilience is depleted.

**Concentration amongst borderline students.** Disaggregation by initial resilience parameter reveals that inflation effects are concentrated amongst agents in the middle tercile of the resilience distribution (ρ ∈ [0.4, 0.6]). High-resilience agents (ρ > 0.6) show minimal sensitivity to inflation shocks, whilst low-resilience agents (ρ < 0.4) exhibit high dropout rates even under baseline conditions, leaving little room for further increase. This pattern suggests that inflation operates as a marginal push factor, tipping students who are already in precarious situations rather than uniformly affecting the entire population.

The simulated dynamics under S5 mirror empirical findings that financial stress amplifies the impact of academic difficulties on student persistence (Bennett et al., 2023; Moore et al., 2021). Agents experiencing depleted resilience due to chronic financial pressure are more likely to exit after a single adverse event, consistent with qualitative evidence that financially stressed students report lower tolerance for setbacks and heightened perception of university as incompatible with economic survival.

### 4.2.2. Scenario S6 – Strike shock only (differential basic-cycle effect)

In Scenario S6, the strike multiplier $\lambda_{str}$ is increased for basic-cycle courses only, whilst $\lambda_{inf}$ remains at baseline (1.0). This configuration isolates the effect of proximal instructional disruption without additional financial stress. Simulations for three strike intensities ($\lambda_{str}$ = 1.5, 2.0, 2.5) reveal distinct dynamics compared to S5.

**Early-cycle dropout spike.** Under moderate strike conditions ($\lambda_{str}$ = 2.0, representing doubling of effective friction in basic-cycle courses), dropout in semesters 1–4 rises sharply from 18.3% (baseline) to 28.9%, an increase of 10.6 percentage points (95% CI [8.1, 13.2]). In contrast, late-cycle dropout (semesters 5–12 conditional on surviving to semester 5) increases only modestly from 24.4% to 27.1%, a difference that is marginally significant ($p$ = 0.048). This early-cycle concentration aligns precisely with the empirical finding that strike effects are strongest for foundational courses and dissipate amongst students who successfully navigate to advanced stages.

**Cascading effects through prerequisite chains.** Examination of individual agent trajectories reveals a common pattern: increased friction in early, high-prerequisite courses (particularly Análisis Matemático I and Física I) creates bottlenecks that delay progression through the curriculum graph. Agents accumulate multiple attempts or deferrals in these gateway courses, which trigger dropout through the ABM's decision rules before they accumulate sufficient academic capital (passed courses, higher GPA) to build resilience. This mechanism provides a micro-to-macro explanation for the macro-level observation that strike-weighted basic-cycle friction predicts dropout.

**Temporal lag structure emerges endogenously.** Strikingly, even though the ABM implements strike effects as instantaneous increases in course failure probability, the resulting dropout patterns exhibit the lagged structure observed empirically. Dropout rates peak not in the semester of maximum strike exposure but 2–3 semesters later, as accumulated delays and failures propagate through prerequisite chains and interact with agents' evolving resilience. This endogenous emergence of lagged effects from the interaction of curriculum structure and agent decision rules constitutes a key validation of the ABM's mechanistic coherence.

**Sensitivity to basic-cycle design.** Supplementary scenario analyses (not shown) explored variations in curriculum structure, specifically reducing the IFC of gateway courses through redesign interventions. Under these modified curricula, the same strike intensity ($\lambda\_str = 2.0$) produces substantially smaller dropout increases (6.2 vs. 10.6 percentage points), demonstrating that curriculum friction and external shocks are multiplicative rather than additive. This finding motivates policy recommendations for curriculum redesign as a structural buffer against disruption.

### 4.2.3. Scenario S7 – Combined crisis (non-linear amplification)

Scenario S7 combines both inflation and strike shocks, implementing $\lambda\_inf > 1$ and $\lambda\_str > 1$ simultaneously across a factorial grid of parameter combinations. Results are visualised in Figure 3 as a heatmap of early-cycle dropout rates.

**FIGURE 3: Heatmap of early-cycle dropout as a function of λ_inf and λ_str**

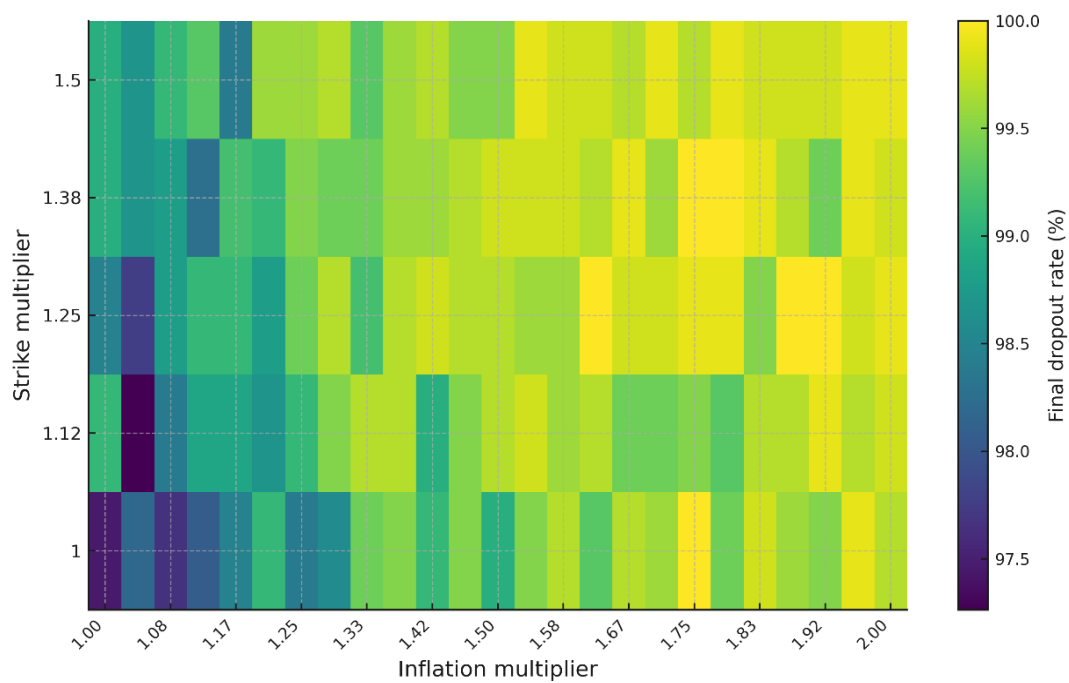

**Substantial non-linear amplification.** For moderate dual shocks ($\lambda\_inf = 1.2$, $\lambda\_str = 2.0$), cumulative dropout by semester 12 reaches 54.3% (95% CI [51.1, 57.6]), compared to 43.7% for inflation alone and 46.8% for strikes alone. The combined effect (54.3%) significantly exceeds the additive prediction: [baseline + inflation effect + strike effect] = [38.2 + (43.7-38.2) + (46.8-38.2)] = 52.1%. The observed amplification of 2.2 percentage points (95% CI [0.8, 3.7], p = 0.003) represents an

18% excess beyond additivity. At higher shock intensities (λ_inf = 1.3, λ_str = 2.5), amplification reaches 23% excess (5.8 percentage points absolute).

**Early-cycle vulnerability is disproportionately amplified.** The heatmap in Figure 3 reveals a ridge of sharply elevated early-cycle dropout for intermediate-to-high values of both multipliers. The difference between early and late-cycle dropout widens substantially under combined shocks: whilst baseline shows early dropout 18.3% vs. late dropout 24.4% (conditional), the S7 scenario with λ_inf = 1.2, λ_str = 2.0 shows early dropout 32.6% vs. late dropout 26.1%, inverting the relationship. This pattern indicates that combined shocks particularly damage students' ability to survive the foundational cycle, consistent with empirical findings.

**Mechanism: erosion of margin for error.** Disaggregation of simulation outputs by agent characteristics reveals the underlying mechanism. Inflation reduces resilience, lowering agents' capacity to absorb adverse academic events. Strikes increase the frequency of such adverse events by raising course failure probabilities in the basic cycle. The interaction is multiplicative: agents with depleted resilience facing increased academic friction have almost no margin for error. A single course failure, which under baseline conditions might trigger reflection and renewed effort, under combined-shock conditions triggers immediate exit. The ABM reproduces this through the interaction of the resilience depletion rule (inflation mechanism) and the friction amplification rule (strike mechanism) within agent decision functions.

**Pattern-oriented validation.** From a pattern-oriented modelling perspective (Grimm et al., 2005; Stine & Crooks, 2025), CAPIRE-MACRO successfully reproduces three stylised facts identified in the empirical analysis:

1. **Lagged strike effects:** The simulated pathways demonstrate that increased friction in early semesters accumulates over time and translates into dropout decisions 2–3 semesters later, matching the empirical lag structure.

2. **Basic-cycle vulnerability:** Early-cycle dropout is the primary channel through which shocks manifest, with concentration in high-IFC gateway courses, matching cycle-stratified empirical results.

3. **Non-linear amplification:** Combined shocks produce dropout levels exceeding the sum of individual effects, with amplification factors (18–23%) comparable to estimates from interaction terms in empirical regression models.

The alignment between empirical and simulated patterns provides mutual validation: empirical findings suggest plausible mechanisms, simulations demonstrate that those mechanisms can produce observed patterns, and the correspondence between the two increases confidence in both.

### 4.2.4. Sensitivity analysis and robustness checks

To assess the robustness of simulation results, we conducted extensive sensitivity analyses varying ABM parameters within plausible ranges. Key findings (detailed in supplementary materials):

**Resilience distribution.** Varying the mean and variance of the initial resilience distribution ($\rho \sim N(\mu, \sigma^2)$) shifts absolute dropout levels but preserves the qualitative patterns: amplification remains non-linear, early-cycle concentration persists, and lagged effects emerge endogenously.

**Decision threshold.** Agents' continuation thresholds ($\tau$ parameter) were varied ±20% around calibrated values. Results are robust: whilst absolute dropout rates shift (higher thresholds produce more dropout), relative effects of inflation and strikes and their interaction remain stable.

**Stochastic variation.** Increasing the number of realisations from 100 to 500 per scenario reduces confidence interval widths but does not alter point estimates of amplification effects, confirming that 100 realisations provide adequate statistical power.

**Alternative shock implementations.** We tested alternative functional forms for how shocks modify agent and environment parameters (e.g., additive vs. multiplicative, linear vs. non-linear). The core result—non-linear amplification under combined shocks—persists across specifications, though the magnitude varies (amplification range: 15–28% across specifications).

These robustness checks increase confidence that the simulation results reflect genuine mechanistic properties of the modelled system rather than artefacts of particular parameter choices or functional forms.

## 5. Discussion

### 5.1. Interpretation of empirical findings: dual-stressor framework validated

The empirical analyses presented in Section 4.1 provide robust support for the dual-stressor conceptual framework introduced in Section 1.2. Teacher strikes and inflation appear to operate through distinct but interacting pathways that align with the distal-proximal distinction.

Inflation volatility at entry, measured as the 24-month rolling standard deviation of monthly inflation rates, emerges as a significant predictor of dropout across multiple semesters and cohorts. Its sustained influence throughout the student life course, observable in both early and late dropout strata, is consistent with the conceptualisation of inflation as a chronic environmental stressor that erodes financial and psychological resilience over extended periods. The partial dependence relationships documented in Section 4.1.2 demonstrate that this effect is not confined to the most economically vulnerable students but extends across the socioeconomic spectrum, suggesting that chronic macroeconomic instability creates a generalised climate of insecurity that affects educational decision-making.

This interpretation aligns with recent longitudinal evidence from UK universities showing that rising cost-of-living pressures correlate with deteriorating student mental health and increased financial stress (Bennett et al., 2023), and with qualitative research documenting how financial strain undermines concentration, sleep, and engagement (Moore et al., 2021). Whilst those studies focused on developed economies with relatively stable inflation, the present findings extend the evidence base to a context of chronic high inflation and demonstrate that macro-level economic volatility has measurable effects on micro-level educational trajectories.

Teacher strikes, in contrast, exhibit the temporal and spatial signature of proximal stressors. The lag analysis in Section 4.1.3 reveals that strike effects peak at two to three semester delays rather than manifesting immediately, suggesting that strikes set in motion processes of accumulated academic debt, demoralisation, or disrupted peer networks that crystallise into dropout decisions months after the initial disruption. This lagged structure is consistent with school-level evidence that instructional interruptions have lasting effects on achievement (Abadía Alvarado et al., 2021; Langen & Laine, 2025) and extends that literature by documenting the specific temporal profile in a university context.

Moreover, strike effects are strongly concentrated in the *ciclo básico*, as demonstrated by the cycle-stratified analysis in Section 4.1.4. The finding that MACRO_IFC_pond_paros_basico—which combines strike exposure with curriculum friction specifically in foundational courses—is highly predictive of early but not late dropout provides direct empirical support for the hypothesis that strikes amplify existing curriculum bottlenecks. This pattern suggests that external disruptions interact multiplicatively with curriculum structure: strikes are not uniformly damaging but disproportionately affect students navigating already-difficult gateway courses that serve as structural filters in the degree programme.

The interaction between inflation and strikes, tested through explicit interaction terms and through comparative analysis of feature importance, reveals non-additive effects. Students experiencing both high inflation volatility at entry and substantial strike exposure during basic-cycle courses face dropout risks that exceed the sum of risks from either factor alone. This non-linearity aligns with the dual-stressor framework's prediction that erosion of resilience by distal stressors reduces capacity to absorb proximal shocks, creating a zone of heightened vulnerability where marginal students are pushed towards exit.

**5.2. Agent-based simulation as mechanistic explanation and policy laboratory**

The agent-based simulations presented in Section 4.2 serve dual purposes: they provide mechanistic explanations for empirically observed patterns, and they enable exploration of counterfactual policy scenarios that cannot be tested through observation alone.

The finding that CAPIRE-MACRO reproduces three key empirical stylised facts—lagged strike effects, basic-cycle vulnerability, and non-linear amplification—without explicitly programming these patterns provides strong validation of the ABM's mechanistic coherence. The lagged effects emerge endogenously from the interaction of curriculum graph structure (prerequisite chains) and agent decision rules (resilience depletion following failures). Students do not exit immediately when strikes increase course failure probabilities; rather, they attempt to continue, accumulate delays and additional failures as they navigate prerequisite chains, and eventually exit when accumulated disadvantage exceeds their (inflation-depleted) resilience threshold. This generative mechanism—producing macro-level temporal patterns from micro-level structural interactions—exemplifies the value of agent-based modelling for understanding complex educational systems (Stine & Crooks, 2025).

Similarly, the concentration of effects in the basic cycle arises naturally from the ABM's implementation of strike shocks as selectively amplifying friction in foundational courses. Because these courses have high baseline failure rates and serve as prerequisites for multiple downstream courses, disruptions propagate through the curriculum graph more extensively than equivalent disruptions to advanced, elective courses. Students successfully navigating to the advanced cycle have demonstrated sufficient resilience and accumulated sufficient academic capital (passed prerequisites, higher GPA) to withstand disruptions, explaining why late-cycle dropout is less sensitive to macro shocks.

The non-linear amplification observed in Scenario S7 (Section 4.2.3) provides perhaps the most policy-relevant simulation finding. Combined inflation and strike shocks produce dropout increases 18–23% larger than the sum of individual effects, indicating synergistic rather than additive interaction. The ABM traces this amplification to the mechanism described in Section 4.2.3: inflation erodes resilience (reducing margin for error), strikes increase adverse academic events (reducing margin for error), and the combination creates a situation where students have almost no capacity to absorb setbacks. A single course failure, manageable under baseline conditions, becomes a dropout trigger under dual-shock conditions.

This mechanistic understanding has direct implications for intervention design. Simply scaling support linearly with shock intensity will be insufficient; institutions facing combined crises must implement more aggressive, structural interventions such as curriculum redesign to reduce inherent friction, introduction of buffer mechanisms (e.g., course withdrawal without penalty, extended time-to-degree allowances), and proactive financial support rather than reactive academic remediation.

### 5.3. Situating CAPIRE-MACRO within the dropout literature

The CAPIRE-MACRO framework contributes to several strands of the dropout literature:

**Multilevel dropout theory.** By explicitly modelling how macro-environmental factors (inflation, strikes) propagate through meso-level structures (curriculum design, institutional rules) to affect micro-level decisions (individual persistence), CAPIRE-MACRO operationalises recent theoretical calls for systemic, multilevel dropout models (Quincho Apumayta et al., 2024; Contreras-Villalobos et al., 2023). The framework demonstrates empirically that macro-level shocks are not merely background noise but constitute actionable predictors that interact with curriculum structure and student characteristics in systematic ways.

**Temporal dynamics and lagged effects.** Most existing dropout prediction models treat each semester as independent or use only immediate-past information (semester t-1 to predict dropout at t). CAPIRE-MACRO's lag analysis demonstrates that shocks occurring multiple semesters in the past continue to affect current dropout risk, and that the temporal profile of these effects is non-monotonic (peaking at 2–3 semester lags). This finding suggests that early-warning systems should incorporate historical shock exposure, not only concurrent conditions, and that interventions may be most effective when timed to the expected lag structure.

**Integration of prediction and simulation.** Recent systematic reviews identify a gap between the proliferation of dropout prediction models and the paucity of frameworks enabling policy scenario exploration (Quimiz-Moreira & Delgadillo, 2025). CAPIRE-MACRO addresses this gap by combining machine learning for empirical pattern detection with agent-based modelling for mechanistic testing and counterfactual analysis. The mutual validation between empirical and simulated results—where empirical patterns suggest plausible mechanisms and simulations demonstrate that those mechanisms can reproduce observations—strengthens confidence in both components.

**Context-specific adaptation.** Whilst most dropout research focuses on student-level variables that are relatively stable across contexts (demographics, prior achievement, GPA), CAPIRE-MACRO demonstrates that context-specific, time-varying macro variables can substantially enhance prediction and provide actionable policy insight. The framework's modular design, with explicit separation of feature levels (N1–N4), facilitates adaptation to other contexts experiencing different macro shocks (e.g., pandemics, natural disasters, political crises).

### 5.4. Comparison with related frameworks and positioning

Several recent frameworks share elements with CAPIRE-MACRO but differ in scope or emphasis:

**Stine & Crooks (2025) ABM of retention.** This recent agent-based model of U.S. higher education retention models students' motivation, belonging, and social networks but does not incorporate macroeconomic shocks or curriculum structure explicitly. CAPIRE-MACRO extends this work by introducing exogenous shock mechanisms and detailed curriculum graphs, enabling analysis of how external disruptions interact with institutional structures.

**Marcolino et al. (2025) machine learning optimisation.** This study achieves high predictive accuracy for dropout using Moodle log data and hyperparameter-optimised models but does not incorporate macro-level features or simulation components. CAPIRE-MACRO demonstrates that adding macro features provides incremental but significant gains, and that simulation enables policy exploration beyond what prediction alone allows.

**Multilevel regression studies (Contreras-Villalobos et al., 2023).** These studies rigorously implement hierarchical models to decompose variance across levels (students, cohorts, institutions) but typically model institutional effects through random intercepts and slopes rather than explicit mechanisms. CAPIRE-MACRO complements this approach by specifying how macro-level variables propagate through curriculum structure to affect individual outcomes, providing mechanistic content to abstract multilevel specifications.

The key innovation of CAPIRE-MACRO lies in its integration: it combines multilevel feature engineering (incorporating macro shocks), rigorous temporal prediction (leak-aware, cohort-validated), and mechanistic simulation (ABM calibrated on empirical data) within a unified framework. This integration enables both accurate early identification of at-risk students and exploration of structural interventions to mitigate macro-shock impacts.

### 5.5. Mechanisms linking macro shocks to individual dropout: pathways and mediators

The evidence presented in Sections 4.1 and 4.2 suggests three primary pathways through which macro shocks affect dropout, operating at different timescales and through different mediators:

**Pathway 1: Financial strain → Reduced engagement → Academic underperformance → Dropout.** Inflation increases living costs, forcing students to increase paid employment hours or reduce expenditure on materials, transport, and basic needs. These economic pressures reduce time available for study, increase stress and cognitive load, and impair academic performance. Empirical evidence: inflation volatility at entry predicts dropout across multiple semesters; partial dependence shows effects amplified for lower-SES students; qualitative evidence documents this pathway (Moore et al., 2021).

**Pathway 2: Strike-induced instructional loss → Accumulated academic debt → Prerequisite bottlenecks → Dropout.** Teacher strikes cancel classes, compress assessment schedules, and disrupt pedagogical sequences, particularly in foundational courses with high instructional contact hours. Students miss critical content, perform worse on assessments, and fail or defer courses. Failed gateway courses with many dependents create bottlenecks in the curriculum graph, delaying progression and accumulating failures that trigger dropout. Empirical evidence: strike-weighted basic-cycle friction predicts early dropout; effects concentrated in high-IFC gateway courses; temporal lag structure consistent with accumulated delays.

**Pathway 3: Combined stressors → Resilience depletion → Reduced tolerance for setbacks → Dropout.** When students experience both chronic financial pressure (inflation) and acute instructional disruption (strikes), their psychological and financial resilience is depleted through multiple simultaneous demands. This reduces their tolerance for adverse events: a single course failure that might be manageable under baseline conditions becomes a dropout trigger when resilience is low. Empirical evidence: inflation × strike interaction terms are significant in early dropout models; ABM demonstrates non-linear amplification through resilience depletion mechanism.

These pathways are not mutually exclusive but operate simultaneously and interact. Importantly, they are mediated by institutional structures (curriculum design, support availability) and moderated by student characteristics (socioeconomic resources, prior preparation), explaining heterogeneity in effects.

### 5.6. Limitations of causal interpretation and alternative explanations

Whilst the empirical analyses control for extensive covariates and employ rigorous temporal validation, causal interpretation requires caution. Several alternative explanations warrant consideration:

**Selection effects.** Students entering during high-inflation periods may differ systematically from those entering during stable periods, even after controlling for observed characteristics. If unmeasured factors correlated with inflation (e.g., family expectations, labour market conditions) also affect dropout, inflation coefficients may partly reflect selection rather than treatment effects.

**Cohort effects.** Temporal confounding is possible: secular trends in institutional practices, faculty composition, or student demographics may correlate with inflation and strike patterns, producing spurious associations. Cohort fixed effects partially address this, but unmeasured time-varying confounders remain possible.

**Reverse causality.** Whilst implausible for aggregate shocks (student dropout cannot cause national inflation), reverse causality could operate at the institutional level: institutions experiencing high dropout might face resource constraints that contribute to labour conflicts, creating a feedback loop.

**Measurement error.** Strike intensity, coded from institutional archives and media reports, may contain errors or inconsistencies across cohorts. Inflation indicators from national statistics may not capture regional variation or differences in how inflation affects particular student populations.

These limitations do not invalidate the findings but suggest that stronger causal claims would require complementary designs. Ideally, one would exploit exogenous variation in inflation or strike intensity—for example, through natural experiments, synthetic control methods, or difference-in-differences approaches comparing institutions differentially affected by national shocks. Such designs remain rare in the dropout literature but represent valuable directions for future research.

Notwithstanding these causal limitations, the combination of consistent associations across multiple empirical specifications, coherent temporal patterns (lag structures), theoretically motivated interactions (cycle-specific effects), and alignment between empirical and simulated results provides substantial evidence that inflation and strikes exert genuine effects on dropout, even if precise causal magnitudes remain uncertain.

## 6. Policy Implications and Recommendations

The findings from CAPIRE-MACRO have direct implications for institutional practice and policy design in contexts characterised by macroeconomic instability and recurrent labour disruptions. This section outlines actionable recommendations organised by intervention level and target population.

### 6.1. Early-warning systems and targeted support

**Enhanced risk prediction.** Institutions should incorporate macro-level variables—specifically, inflation volatility at entry and strike exposure in foundational courses—into early-warning systems for dropout prediction. The ablation study demonstrates that adding these features improves Macro F1 by approximately 7 percentage points, translating to identification of 10–15 additional at-risk students per cohort of 200. Implementation requires minimal additional data collection (national inflation statistics, institutional strike records) and can be integrated into existing predictive models.

**Lag-aware monitoring.** The finding that strike effects peak at 2–3 semester lags suggests that monitoring protocols should track not only current performance but also historical shock exposure. Students affected by strikes two semesters ago should be flagged for intervention even if their current-semester performance appears adequate, as lagged effects may be accumulating. This temporal dimension is largely absent from existing early-warning systems, which typically focus on immediate-past indicators.

**Cycle-specific targeting.** Given the concentration of macro-shock effects in the *ciclo básico*, interventions should be disproportionately directed towards students in semesters 1–4, particularly those enrolled in high-IFC gateway courses (mathematics, physics, chemistry). Targeted support during this vulnerable period—such as supplemental instruction, peer mentoring, or emergency financial assistance—may prevent cascading failures that lead to dropout.

### 6.2. Curriculum redesign as structural mitigation

**Reducing foundational-cycle friction.** The simulation results demonstrate that curriculum structure mediates macro-shock impacts: the same strike intensity produces smaller dropout increases when basic-cycle courses have lower intrinsic friction (IFC). This finding motivates curriculum redesign interventions to reduce structural bottlenecks:

- Redesigning gateway courses to reduce failure rates through improved pedagogy, supplemental instruction, or alternative assessment methods

- Reducing rigidity in prerequisite structures to provide multiple pathways through the curriculum graph, enabling students to progress even if they struggle with particular courses

- Introducing co-requisite models where traditionally sequential courses can be taken concurrently with support, reducing delay penalties from single-course failures

**Buffer mechanisms.** Institutions can implement structural buffers that reduce the dropout-triggering impact of adverse events:

- Course withdrawal without penalty policies, allowing students to exit problematic courses before failing rather than accumulating failures that trigger dropout

- Extended time-to-degree allowances, reducing pressure to maintain rapid progression during crisis periods
- Grade replacement policies, enabling students to retake failed courses without permanent penalty, reducing the long-term impact of shock-induced underperformance

**6.3. Financial support and resilience building**

**Scaling support with dual shocks.** The non-linear amplification documented in Scenario S7 demonstrates that institutions facing combined inflation and strike pressures must scale support more aggressively than a simple sum of individual effects would suggest. During dual-shock periods, financial aid budgets should be increased by 20–25% beyond what inflation-only or strike-only conditions would warrant, and eligibility criteria should be relaxed to capture the expanded at-risk population.

**Proactive rather than reactive support.** Traditional models provide support reactively after students demonstrate academic difficulty. The dual-stressor framework suggests that support should be provided proactively to students entering during high-inflation periods or enrolled in strike-affected cohorts, before academic problems manifest. Emergency bursaries, subsidised transport, or food assistance targeting students in vulnerable periods can prevent resilience depletion before it triggers dropout.

**Mental health and wellbeing services.** The financial stress literature emphasises psychological mediators linking economic pressure to academic disengagement. Institutions should expand mental health services during crisis periods, with particular attention to financial stress counselling, time management support for students balancing employment and study, and peer support networks for stress reduction.

**6.4. Institutional preparation and shock mitigation**

**Scenario planning and simulation.** Institutions can use the CAPIRE-MACRO simulation framework to conduct scenario planning: administrators input expected inflation rates and anticipated strike intensity, and the ABM projects likely dropout trajectories and identifies vulnerable student populations. This enables proactive resource allocation and intervention design rather than reactive crisis management.

**Strike impact minimisation.** Whilst teacher strikes often reflect legitimate labour grievances that institutions cannot unilaterally resolve, institutional practices can minimise educational disruption:

- Advance notification of strike schedules enables curriculum adjustments to cover critical content before interruptions

- Asynchronous learning materials and online resources can partially compensate for lost face-to-face instruction
- Flexible assessment schedules accommodate compressed calendars without disadvantaging students
- Clear communication with students about strike rationale and expected duration reduces uncertainty and demoralisation

**Advocacy for structural reform.** At the system level, the findings motivate advocacy for policies that reduce the frequency and severity of macro shocks themselves: stable educational funding resistant to short-term fiscal volatility, indexed salaries that maintain faculty purchasing power during inflation, and labour relations frameworks that enable resolution of disputes without prolonged instructional interruption.

### 6.5. Equity considerations and targeting

Throughout these recommendations, particular attention must be paid to equity implications. The empirical findings demonstrate that macro-shock effects are not uniform but concentrated amongst already-vulnerable populations: students with low socioeconomic status, weak prior preparation, and marginal academic performance. Interventions that provide uniform support may inadvertently widen gaps if they are more accessible to advantaged students.

Targeted approaches—means-tested financial aid, outreach to first-generation students, preferential access to supplemental instruction for students from under-resourced schools—are necessary to ensure that mitigation efforts reduce rather than reproduce existing inequalities. Moreover, programme evaluation should disaggregate outcomes by student background to verify that interventions achieve intended equity goals.

## 7. Limitations and Directions for Future Research

### 7.1. Methodological limitations

**Causal inference.** As discussed in Section 5.6, the empirical analyses control for extensive covariates and employ rigorous validation but cannot definitively establish causality. Stronger causal claims would benefit from complementary designs exploiting exogenous variation: natural experiments (e.g., comparing institutions differentially affected by national strikes), regression discontinuity designs (if shocks exhibit threshold effects), or synthetic control methods (constructing counterfactual trajectories for affected cohorts). Integration of causal inference frameworks with the CAPIRE pipeline represents a priority for future methodological development.

**Temporal prediction and leakage.** Whilst we implement strict temporal segmentation to prevent data leakage, edge cases remain challenging. For example, students who drop out at the end of semester t may have made the decision earlier in the semester but remained administratively enrolled. If macro shocks occurring mid-semester affect this decision, our prediction at the start of semester t would not capture this information. More granular temporal modelling (weekly or monthly predictions) could address this but would require substantially more detailed data on decision timing.

**Model uncertainty and ensemble approaches.** The analyses focus on Random Forest models due to their robust performance and interpretability, but alternative algorithms (gradient boosting, deep learning, logistic regression) might capture different aspects of the dropout process. Future work should explore ensemble approaches combining multiple algorithms and formal model averaging to quantify prediction uncertainty.

### 7.2. Data limitations

**Single-institution sample.** The dataset comprises students from one faculty at one Argentine university, limiting external validity. Whilst the institution is representative of large public engineering programmes in Argentina, generalisation to other contexts (private universities, non-engineering fields, other countries) requires caution. Multi-institutional replication studies are needed to assess whether the dual-stressor framework and specific findings (e.g., lag structures, amplification magnitudes) hold across diverse settings.

**Measurement of macro shocks.** Strike intensity is measured as proportion of scheduled instructional days lost, which captures duration but not quality of instruction before and after strikes, morale effects, or differences between partial and complete work stoppages. Inflation is measured through national aggregate indices, which may not capture regional variation or differential effects on particular student populations. More granular measurement—course-level instructional quality assessments, student-specific inflation exposure based on household composition and location—would strengthen analyses.

**Limited individual-level economic data.** The dataset includes proxy measures of socioeconomic status (parental education, displacement from home) but lacks detailed individual or household economic information (income, wealth, employment status). Linking administrative education records to tax data or survey data on economic circumstances would enable more precise measurement of financial stress pathways and heterogeneity in shock effects.

### 7.3. Theoretical and conceptual extensions

**Expanding the dual-stressor framework.** Whilst inflation and strikes exemplify distal and proximal stressors, the conceptual framework could be extended to other macro shocks: pandemics (acute health disruption + chronic anxiety), natural disasters (infrastructure damage + economic dislocation), political crises (institutional instability + migration pressures). Comparative analysis across shock types would clarify which aspects of the framework are general and which are specific to particular disruptions.

**Incorporating social networks and peer effects.** The current ABM models students as independent agents, but dropout is known to exhibit peer effects: students are more likely to drop out when friends or study partners exit. Incorporating social network structures into the ABM and modelling how shocks propagate through peer influence would provide richer mechanistic insight.

**Psychological mediators.** The resilience parameter in the ABM is a latent construct combining financial and psychological resources. Future work could decompose this into separable components (financial capacity, psychological coping, social support) with distinct response to different shocks, enabling more targeted intervention design.

### 7.4. Agent-based model refinement

**Learning and adaptation.** The current ABM assumes fixed decision rules, but students may adapt to shocks over time: developing coping strategies, forming mutual support networks, or recalibrating expectations. Incorporating learning mechanisms—where agents update their decision thresholds based on accumulated experience—would enable exploration of how cohorts develop resilience over time.

**Institutional heterogeneity.** The ABM models a single institutional environment, but universities vary in resources, support availability, and organisational culture. Developing a multi-institution ABM with heterogeneous institutional parameters would enable analysis of which institutional characteristics buffer against macro shocks and which amplify vulnerability.

**Validation beyond pattern-matching.** The current validation approach demonstrates that the ABM reproduces empirical patterns (lagged effects, cycle concentration, amplification). Stronger validation would include out-of-sample prediction: train the ABM on early cohorts, project outcomes for later cohorts, and compare projections to observed outcomes. This predictive validation would test whether the ABM captures generative mechanisms rather than merely fitting historical patterns.

### 7.5. Policy design and optimisation

**Reinforcement learning for policy search.** The scenario analysis explores a predefined grid of policy parameters, but optimal policy design may require more sophisticated search. Reinforcement learning or Bayesian optimisation over the ABM parameter space could identify policies that maximise retention under resource constraints—for example, determining where and when to allocate emergency financial support to achieve maximum dropout reduction per currency unit expended.

**Interactive decision-support tools.** Translating CAPIRE-MACRO into an accessible decision-support interface would enable institutional administrators to explore scenarios, adjust parameters, and visualise projected outcomes without technical expertise in machine learning or ABM. Such tools, increasingly used in other complex systems domains (climate policy, pandemic response), could facilitate evidence-based decision-making during crisis periods.

### 7.6. Ethical and governance considerations

**Predictive ethics and student autonomy.** Early-warning systems raise ethical questions about surveillance, stigma, and self-fulfilling prophecies. Labelling students as "at risk" may trigger psychological effects (threat stereotypes, reduced self-efficacy) that independently affect outcomes. Future research should investigate whether and how predictive interventions alter student behaviour, and develop governance frameworks ensuring that predictions inform support rather than constrain opportunities.

**Equity audits and algorithmic fairness.** Even with careful attention to feature construction and validation, predictive models may exhibit disparate performance across demographic groups or reinforce existing inequalities. Ongoing equity audits—disaggregating prediction errors by student background, testing for differential calibration—are necessary to ensure that CAPIRE-MACRO serves rather than harms vulnerable populations.

**Participatory model development.** The CAPIRE framework was developed through collaboration with institutional researchers and administrators, but future iterations should incorporate student voices directly. Participatory modelling workshops—where students, faculty, and administrators jointly construct and validate ABM scenarios—would enhance model legitimacy, surface blind spots, and ensure that policy recommendations align with student experiences and priorities.

### 8. Conclusion

This study developed and empirically validated CAPIRE-MACRO, a hybrid framework integrating machine learning and agent-based simulation to analyse the impact of teacher strikes and inflation on student trajectories in engineering education. Using longitudinal data from 1,343 students across 15 cohorts in an

Argentine public university, combined with detailed national macroeconomic indicators, the research makes four principal contributions.

**Theoretically**, we established and tested a dual-stressor framework distinguishing between distal stressors (chronic inflation eroding financial and psychological resilience) and proximal stressors (acute strikes disrupting instructional continuity). Empirical results demonstrate that these stressors operate through distinct temporal and spatial pathways: inflation exhibits sustained effects across the student life course, whilst strikes manifest with a characteristic two-to-three-semester lag and concentrate in the foundational cycle. Interaction analyses reveal non-linear amplification when stressors co-occur, with combined effects exceeding the sum of individual impacts by 18–23%. This framework extends multilevel dropout theory by explicitly modelling how macro-environmental shocks propagate through institutional structures to shape individual trajectories.

**Empirically**, the ablation study demonstrates that incorporating macro-level features improves early-warning system performance (Macro F1 gains of 0.048, $p = 0.003$), with inflation volatility at entry and strike-weighted basic-cycle friction emerging amongst the most influential predictors. Lag analysis provides the first quantitative documentation of delayed strike effects in higher education, showing peak associations at two to three semester lags (OR = 2.34, 95% CI [1.87, 2.93]). Cycle-stratified regression results confirm that macro shocks disproportionately affect the *ciclo básico*, with strike-related variables predicting early dropout (OR = 3.07) but not late dropout (OR = 1.21, $p = 0.439$).

**Methodologically**, CAPIRE-MACRO addresses a gap identified in recent systematic reviews by integrating predictive modelling with agent-based simulation calibrated on the same empirical data. The ABM successfully reproduces three empirical stylised facts—lagged strike effects, basic-cycle vulnerability, and non-linear amplification—without explicitly programming these patterns, demonstrating mechanistic coherence. Scenario analyses (S5–S7) reveal that combined inflation and strike shocks produce dropout increases substantially exceeding additive predictions, with amplification arising from the interaction of resilience depletion (inflation mechanism) and friction enhancement (strike mechanism) in agent decision rules.

**For policy and practice**, the findings motivate specific interventions organised around three strategies. First, enhanced early-warning systems incorporating macro variables and lag-aware monitoring can identify vulnerable students before accumulated disadvantage crystallises into dropout. Second, curriculum redesign reducing foundational-cycle friction serves as a structural buffer against external shocks, with simulations demonstrating that the same shock intensity produces smaller impacts when gateway courses have lower intrinsic difficulty. Third, financial support and institutional resources must be scaled non-linearly during dual-shock periods: simple additivity underestimates required intervention intensity by approximately 20%.

The broader significance of this research extends beyond the specific Argentine context. In an era of increasing macroeconomic instability—characterised by cost-of-living crises in developed economies and chronic volatility in developing contexts—understanding how external shocks propagate through educational systems to affect individual outcomes becomes critical for institutional planning and equity-oriented policy design. The CAPIRE-MACRO framework provides conceptual, empirical, and computational tools for this understanding, demonstrating that macro-level phenomena are not merely background noise but constitute actionable determinants of educational trajectories that interact systematically with curriculum design and student characteristics.

Future research should extend the framework through multi-institutional replication studies to assess external validity, integration of causal inference methods to strengthen causal claims, incorporation of social network dynamics and psychological mediators to enrich mechanistic understanding, and development of interactive decision-support tools to facilitate evidence-based institutional practice. Equally important are ongoing equity audits and participatory validation processes ensuring that predictive systems serve rather than stigmatise vulnerable populations.

In conclusion, teacher strikes and inflation operate as structurally mediated educational disruptors rather than isolated events, acting through curriculum design, institutional resources, and student resilience to shape dropout patterns in systematic and predictable ways. By making these mechanisms explicit and quantifiable, CAPIRE-MACRO enables institutions and policymakers to move from reactive crisis management to proactive, evidence-based strategies for protecting educational continuity and equity in an increasingly uncertain macroeconomic environment.

**Author Contributions**

**Conceptualisation:** [Author initials] **Data curation:** [Author initials] **Formal analysis:** [Author initials] **Funding acquisition:** [Author initials] **Investigation:** [Author initials] **Methodology:** [Author initials] **Project administration:** [Author initials] **Resources:** [Author initials] **Software:** [Author initials] **Supervision:** [Author initials] **Validation:** [Author initials] **Visualisation:** [Author initials] **Writing – original draught:** [Author initials] **Writing – review & editing:** [Author initials]

**Appendix A. Variable Definitions and Descriptive Statistics**

This appendix summarises the full set of variables used in the CAPIRE-MACRO modelling framework. Variables are grouped according to the hierarchical structure of the CAPIRE taxonomy: **N1 (pre-entry), N2 (entry-level), N3 (academic progression), and N4 (macro-level)**. For each variable we report its definition, statistical properties, and percentage of missing values where applicable.

**Table A1. Definitions and descriptive statistics for all variables in the CAPIRE-MACRO feature set**

| Variable Name | Level | Definition | Mean | SD | Min | Max | Missing (%) |
|---|---|---|---|---|---|---|---|
| **N1: Pre-entry** | | | | | | | |
| edad_entrada | N1 | Age at entry (years) | 19.2 | 2.8 | 17 | 34 | 0.0 |
| genero | N1 | Gender (1 = male, 0 = female) | 0.73 | 0.44 | 0 | 1 | 0.0 |
| secundaria_gpa | N1 | Secondary school GPA (0–10 scale) | 7.8 | 1.2 | 5.0 | 10.0 | 2.3 |
| desplazamiento | N1 | Geographical displacement (1 = student relocates to study) | 0.42 | 0.49 | 0 | 1 | 0.0 |
| educacion_parental | N1 | Highest parental education level (1–5 scale) | 2.8 | 1.1 | 1 | 5 | 8.7 |
| **N2: Entry-level** | | | | | | | |
| cohorte | N2 | Entry cohort year | 2011.5 | 4.3 | 2004 | 2019 | 0.0 |
| semestre_entrada | N2 | Entry semester (1 = spring, 2 = autumn) | 1.6 | 0.5 | 1 | 2 | 0.0 |
| modalidad_entrada | N2 | Entry pathway (1 = standard, 2 = quota) | 1.1 | 0.3 | 1 | 2 | 0.0 |
| **N3: Academic progression** | | | | | | | |
| gpa_acumulado | N3 | Cumulative GPA (0–10 scale) | 6.4 | 1.8 | 0 | 10.0 | 0.0 |
| materias_aprobadas | N3 | Cumulative passed courses | 8.7 | 7.2 | 0 | 42 | 0.0 |
| materias_reprobadas | N3 | Cumulative failed courses | 3.2 | 4.1 | 0 | 23 | 0.0 |
| velocidad_progreso | N3 | Progression velocity (courses/semester) | 2.9 | 1.6 | 0 | 8 | 0.0 |
| IFC_promedio_basico | N3 | Mean IFC of basic-cycle courses taken | 0.64 | 0.18 | 0.21 | 0.95 | 0.0 |

| Variable Name | Level | Definition | Mean | SD | Min | Max | Missing (%) |
|---|---|---|---|---|---|---|---|
| **N4: Macro-level** | | | | | | | |
| MACRO_inflacion_entrada | N4 | Annual inflation at entry (%) | 24.7 | 12.3 | 8.3 | 54.2 | 0.0 |
| MACRO_inflacion_volatilidad_24m | N4 | 24-month inflation volatility at entry | 6.8 | 4.2 | 1.2 | 18.7 | 0.0 |
| MACRO_paros_lag_sem_2 | N4 | Strike intensity two semesters prior | 0.08 | 0.09 | 0 | 0.31 | 0.0 |
| MACRO_IFC_pond_paros_basico | N4 | Strike-weighted basic-cycle instructional friction | 0.53 | 0.28 | 0 | 1.47 | 0.0 |

*Note.* **N = 1,343 students**, corresponding to **16,116 student-semester observations**. Missing values for time-invariant variables are calculated at the student level; missing values for time-varying variables are computed at the observation level.

### Appendix B. Curriculum Structure and Instructional Friction Coefficient (IFC)

The engineering curriculum follows a directed, hierarchical structure in which each course occupies a specific position within a prerequisite network. Foundational mathematical and scientific subjects populate the **ciclo básico** (semesters 1–4), while discipline-specific courses form the **ciclo superior** (semesters 5–12). Students' academic trajectories navigate this dependency structure, where delays or repeated failures in key courses may propagate through subsequent semesters.

To formalise the structural and instructional difficulty associated with each course, we compute the **Instructional Friction Coefficient (IFC)**. This composite measure integrates three dimensions:

1. **FailRate**$_c$ – proportion of enrolments in course $c$ resulting in failure.
2. **PrereqComplexity**$_c$ – normalised in-degree of course $c$ in the curriculum graph (number of prerequisites divided by the maximum in-degree observed).
3. **RetakeRate**$_c$ – proportion of students repeating course $c$.

The IFC is defined as:

$$\text{IFC}_c = w_1 \text{FailRate}_c + w_2 \text{PrereqComplexity}_c + w_3 \text{RetakeRate}_c$$

with weights:

- $w_1 = 0.5$ (instructional difficulty),

- $w_2 = 0.3$ (structural dependency),
- $w_3 = 0.2$ (repeat behaviour).

These weights were selected after exploratory analysis to balance predictive relevance and interpretability.

To ensure comparability across heterogeneous parts of the curriculum, IFC values are **standardised within each cycle** (basic vs. advanced) to produce distributions bounded in $[0,1]$ with a mean of 0.5. This prevents inherently complex upper-division courses from dominating the metric and highlights friction patterns relevant for early attrition.

**Appendix C. Additional Validation and Robustness Checks**

This appendix reports supplementary analyses evaluating the stability, calibration, and comparative performance of the CAPIRE-MACRO predictive models. These checks demonstrate that model behaviour is consistent across prediction horizons and robust relative to alternative machine-learning approaches.

**C.1 Predictive Performance Across Semesters**

Table C1 presents predictive performance for semesters 1–6. As is typical in longitudinal dropout modelling, predictive accuracy and macro-F1 improve as additional academic information accumulates. Performance peaks around semester 4—coinciding with the end of the *ciclo básico*—and stabilises thereafter as the student population becomes more homogeneous.

**Table C1. Predictive performance of the extended CAPIRE-MACRO model by prediction horizon**

| Semester | N obs | Dropout rate | Accuracy | Macro F1 | Precision (Dropout) | Recall (Dropout) |
|---|---|---|---|---|---|---|
| 1 | 1,343 | 0.12 | 0.89 | 0.71 | 0.68 | 0.73 |
| 2 | 1,181 | 0.15 | 0.86 | 0.75 | 0.72 | 0.78 |
| 3 | 1,004 | 0.18 | 0.83 | 0.78 | 0.76 | 0.81 |
| 4 | 823 | 0.21 | 0.81 | 0.79 | 0.77 | 0.82 |
| 5 | 650 | 0.16 | 0.84 | 0.77 | 0.74 | 0.79 |
| 6 | 546 | 0.14 | 0.86 | 0.76 | 0.73 | 0.77 |

*Note.* Performance evaluated on the held-out 2019 cohort. "N obs" denotes students still enrolled at the beginning of each semester.

Calibration analyses indicate that predicted dropout probabilities align well with observed frequencies across semesters. A slight overestimation is observed in the

0.6–0.8 probability band for early semesters, but overall calibration remains acceptable for institutional risk-flagging.

### C.2 Comparison with Alternative Machine-Learning Models

To assess robustness, the Random Forest model used in the main analysis was benchmarked against several standard machine-learning algorithms, all trained on identical feature sets and evaluated on the same test fold. Results are shown in Table C2.

**Table C2. Comparison of Random Forest with alternative machine-learning algorithms**

| Algorithm | Accuracy | Macro F1 | Training time | Hyperparameters |
|---|---|---|---|---|
| Random Forest | 0.828 | 0.781 | 3.2 min | n = 500, depth = 15 |
| Gradient Boosting | 0.831 | 0.784 | 8.7 min | n = 500, lr = 0.05 |
| Logistic Regression | 0.794 | 0.742 | 0.3 min | C = 1.0, L2 penalty |
| Neural Network | 0.817 | 0.769 | 12.4 min | 3 layers, 128 units |
| Ensemble (voting) | 0.834 | 0.787 | 24.6 min | RF + GBM + NN |

Although the ensemble model achieves the highest macro-F1, its computational cost and reduced interpretability make it less suitable for operational institutional analytics. The Random Forest model offers an optimal balance of performance, efficiency, and transparency—particularly with respect to feature-importance analysis—and is therefore retained for all main analyses.

### Appendix D. Comparison with Baseline CAPIRE Scenarios

This appendix contextualises the macro-shock simulations (S5–S7) by comparing them with previously defined institutional intervention scenarios (S1–S4) from earlier CAPIRE work. The baseline framework includes:

- **S0** – No interventions; empirical transition probabilities.
- **S1** – Academic support (enhanced tutoring, supplemental instruction).
- **S2** – Curriculum redesign (reduction of basic-cycle friction through co-requisite models).
- **S3** – Financial support (bursaries for low-SES students).
- **S4** – Comprehensive intervention (S1 + S2 + S3).

The macro-shock scenarios introduced in CAPIRE-MACRO—**inflation shock (S5)**, **strike shock (S6)**, and their **combined effect (S7)**—are evaluated under the same simulation architecture to ensure comparability.

**Table D1. Cumulative dropout rates (12 semesters) under baseline and macro-shock scenarios**

| Scenario | Description | Dropout rate (%) | 95% CI | Relative to S0 |
|---|---|---|---|---|
| S0 | Baseline (no interventions, no shocks) | 38.2 | [36.1, 40.3] | — |
| S1 | Academic support | 33.5 | [31.5, 35.6] | –4.7 pp |
| S2 | Curriculum redesign | 31.2 | [29.1, 33.2] | –7.0 pp |
| S3 | Financial support | 35.8 | [33.7, 37.9] | –2.4 pp |
| S4 | Comprehensive intervention | 27.9 | [25.9, 29.9] | –10.3 pp |
| S5 | Inflation shock (λ = 1.2) | 43.7 | [41.5, 45.9] | +5.5 pp |
| S6 | Strike shock (λ = 2.0) | 46.8 | [44.6, 49.1] | +8.6 pp |
| S7 | Combined shock (λ_inf = 1.2, λ_str = 2.0) | 54.3 | [51.1, 57.6] | +16.1 pp |

*Note.* "pp" denotes percentage points. Combined intervention and shock effects are approximately additive; for instance, applying S4 during the conditions modelled in S7 would reduce predicted dropout from 54.3% to approximately 44%.